\batchmode
\documentclass{aa}  
\usepackage{graphicx}
\usepackage{natbib}
\bibpunct{(}{)}{;}{a}{}{,}
\usepackage{txfonts}
\begin{document}

   \titlerunning{Sunspot areas and tilt angles for solar cycles 7--10}
   \authorrunning{V. Senthamizh Pavai et al.}

   \title{Sunspot areas and tilt angles for solar cycles 7--10}

   \author{V. Senthamizh Pavai\inst{1}
          \and
          R. Arlt\inst{1}
          \and
          M. Dasi-Espuig\inst{3}
          \and
          N.A. Krivova\inst{2}
          \and
          S.K. Solanki\inst{2, 4}
          }

   \institute{Leibniz-Institut f\"ur Astrophysik Potsdam, An der Sternwarte 16, 14482 Potsdam, Germany\\
             \email{svalliappan@aip.de, rarlt@aip.de}
             \and
             Max-Planck-Institut f\"ur Sonnensystemforschung, Justus-von-Liebig-Weg 3, 37077 G\"ottingen, Germany
             \and
             Imperial College London, Blackett Laboratory, Prince Consort Road, London SW7 2AZ, UK 
             \and
             School of Space Research, Kyung Hee University, Yongin, 446-101 Gyeonggi, Republic of Korea}

   \date{Received 2015; accepted 2015}

 
  \abstract
{}
  {Extending the knowledge about the properties of solar cycles 
   into the past is essential for understanding the solar dynamo. 
   This paper aims at estimating areas of sunspots 
   observed by Schwabe in 1825--1867 and at calculating the tilt 
   angles of sunspot groups.}
   {The sunspot sizes in Schwabe's drawings are not to scale and 
   need to be converted into physical sunspot areas. We employed 
   a statistical approach assuming that the area distribution of 
   sunspots was the same in the 19th century as it was in the 20th 
   century.
}
   {Umbral areas for about 130,000 sunspots observed by Schwabe 
   were obtained, as well as the tilt angles of sunspot groups 
   assuming them to be bipolar. There is, of course, no 
   polarity information in the observations. The annually averaged 
   sunspot areas correlate reasonably with sunspot number. We
   derived an average tilt angle by attempting to exclude unipolar
   groups with a minimum separation of the two alleged polarities
   and an outlier rejection method which follows the evolution of
   each group and detects the moment it turns unipolar at its
   decay. As a result, the 
   tilt angles, although displaying considerable scatter, place 
   the leading polarity on average $5\fdg85\pm0\fdg25$ closer to the equator, 
   in good agreement with tilt angles obtained from 20th-century data 
   sets. Sources of uncertainties in the tilt angle determination are
   discussed and need to be addressed whenever different data sets
   are combined. The sunspot area and tilt angle data are provided online. 
}
   {}

   \maketitle

%

\errorstopmode
\section{Introduction}

Solar activity is apparently driven by internal magnetic fields, which 
are roughly oscillatory in time. Sunspots are the most obvious manifestations
of solar activity in visible light, and it was Samuel Heinrich \cite{schwabe44}
who first published a paper on the abundance of sunspots as a cyclic phenomenon.

Apart from the number of sunspots and the various indices that can be
defined from their appearance, there are other properties that contain
information on the underlying process of generating variable magnetic
fields in the solar interior. The most prominent feature is the
distribution of spots in latitude versus time \citep[butterfly diagram;][]{carrington63}.
The latitudes of the spots give us an idea of the location of
the underlying magnetic fields. In the majority of attempts to 
explain the dynamo process of the Sun, it is assumed that it is 
strong azimuthal magnetic fields which emerge as sunspots at the 
solar surface \citep[for a review, see][]{charbonneau2010}. These
internal horizontal (i.e.\ parallel to the solar surface) fields become locally
unstable and form loops eventually penetrating the surface of the Sun.
At this stage, two polarities are formed, which are actually measured
and often accompanied by sunspot groups \citep[for a review, see][]{fan2004}. 
Alternatively, sunspots may form as a consequence of a large-scale
magnetic field suppressing the convective motions and thereby reducing the
turbulent pressure. The lower pressure at the field location compresses
the flux even further leading to further turbulence suppression, and an
instability can occur \citep{kleeorin_ea1989,warnecke_ea2013}.

The radial magnetic field within active regions provides 
poloidal fields to the dynamo system. The production of poloidal 
magnetic flux is an essential ingredient to the sustainability of 
the Sun's large-scale magnetic field. The angle the group polarities 
form with the solar equator is called the tilt angle and was first
measured by \cite{hale_ea1919}. On average, the leading 
polarity of the group is slightly closer to the solar equator 
than the following one. The dependence of the average tilt angle 
on the emergence latitude of the sunspot groups is often referred 
to as Joy's law, according to the paper mentioned above.

Tilt angles from sunspots in white-light images were computed by 
\citet{howard1991} from Mt. Wilson images and by \citet{sivaraman_ea1999}
from Kodaikanal images. From those data, average tilt angles were 
obtained by \citet{dasi_ea2010,dasi_ea2013} for the solar cycles 15--21. An 
anti-correlation between the average tilt angle and the amplitude of
the corresponding cycle was found. Additionally, the product of 
this average and the cycle amplitude correlates significantly with the
strength of the next cycle. We will come back to more recent tilt angle
determinations in Sects.~\ref{averages} and \ref{results}.

Based on magnetograms from Kitt Peak, 
\citet{wang_sheeley1989} determined tilt angles for cycle~21 and 
obtained a large average value of $10\degr$, a result confirmed by 
\citet{stenflo_kosovichev2012} from MDI data. Recently, \citet{wang_ea2015} compared 
tilt angles from white-light images of the Debrecen Photoheliographic 
Database and from Mt.~Wilson magnetograms for cycles~21--23. 
They found that magnetogram tilt angles tend to be larger than 
the ones from sunspot groups in white-light images, both because
a substantial fraction of the white-light tilt angles refer to sunspots
of the same polarity, and because the magnetograms include magnetic 
flux from plage regions typically
showing higher tilts than the sunspots of the same active region. 
We will address the first issue later in this Paper.

This Paper is based on the digitized observations by Samuel Heinrich 
Schwabe \citep{arlt2011} of cycles 7--10 and extends the subsequent 
measurements of all positions and estimates of the sizes of the sunspots
drawn in these manuscripts \citep{arlt_ea2013}. The initial sizes
were in arbitrary units corresponding to pixel areas in the
digital images and may not be to scale. We will describe the
method of converting these sunspot size estimates into physical
areas in Sect.~\ref{areas}, an attempt at defining proper sunspot
groups in Sect.~\ref{groups}, the computations of the tilt angles
in Sect.~\ref{tilts} and will summarize the results in Sect.~\ref{results}.


\begin{figure}
\includegraphics[width=0.49\textwidth]{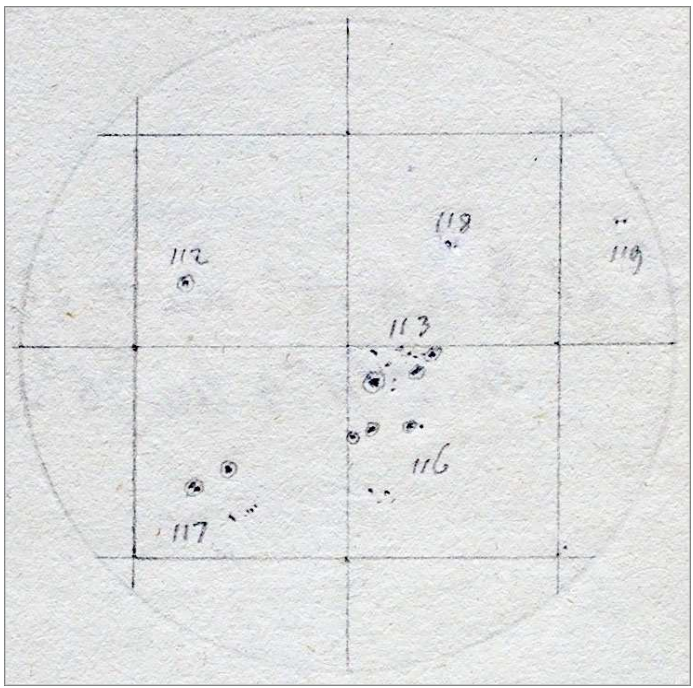}
\caption{Example of the drawing style in the main period of 1831--1867.
This full-disk drawing of 1847 July~22 shows spots with and without
penumbrae. The drawing also shows two group designations (116 and 117) which 
actually refer to two individual groups each (see Sect.~\ref{groups},
also for the treatment of more difficult cases such as group~113).
\label{penumbrae}}
\end{figure}

\begin{figure}
\includegraphics[width=0.49\textwidth]{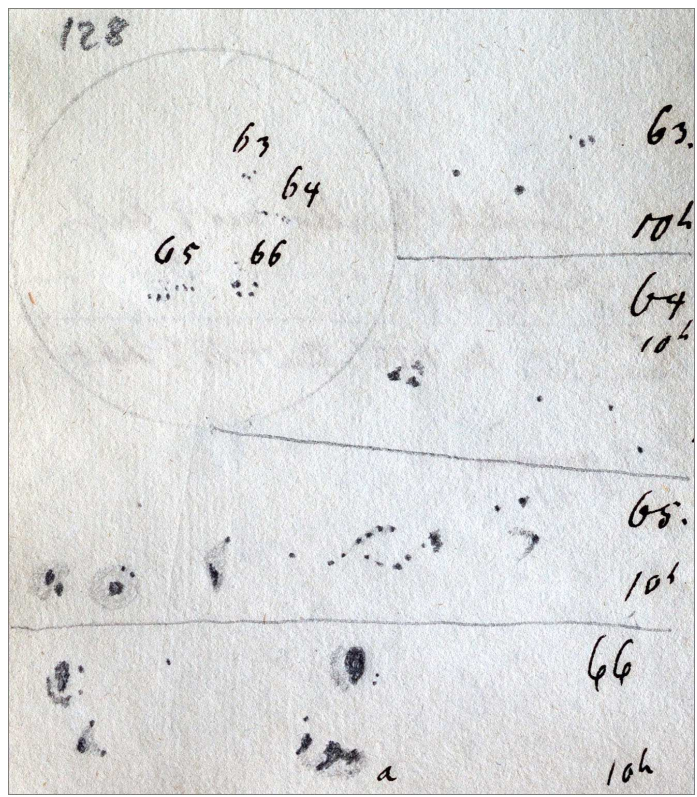}
\caption{Example of the drawing style in the initial period of 1825--1830.
This full-disk drawing of 1827 June~13 shows large spots which combine
several umbrae and at least part of the penumbral area, as is revealed
by the detailed drawings.
\label{clumping}}
\end{figure}

\section{Calibration of sunspot areas}\label{areas}
Apart from the importance to have reliable sunspot area information
for the Schwabe period, we also need to know the individual sunspot areas
for reasonable estimates of the two polarities and their locations in sunspot groups
when determining the tilt angles of sunspot groups. The sunspot
areas may be seen as proxies for the magnetic flux \citep[e.g.]
[for early studies]{houtgast_vansluiters1948,ringnes_jensen1960}, although 
the relation of the two may be complex, as emphasized recently by
\citet{tlatov_pevtsov2014}.

Schwabe plotted the sunspots into relatively small circles of about 5~cm 
diameter, representing the solar disk. Given the finite width of a pencil 
tip, at least small spots must have been plotted with an area larger than
a corresponding structure on the Sun would have at that scale. Pores, if 
plotted to scale, would need to have diameters of 0.04--0.1~mm in such a 
drawing. The umbral areas measured in the drawings therefore need to be
converted into physical areas on the Sun.

There are two ways of obtaining physical sizes of the sunspots drawn 
by Schwabe. The one requires the existence of high-resolution drawings 
by other observers within the observing period of Schwabe for calibration. 
The other is a statistical approach using data sets of the 20th 
century to calibrate the sizes. We will first describe the latter
method, since the number of high-resolution drawings by other observers
that can be employed for the first method is very limited. The statistical approach also
required a splitting of the data into two sets: 1825--1830 and
1831--1867. This is due to a change in the drawing style after 1831 Jan 1,
as demonstrated in Figs.~\ref{penumbrae} and \ref{clumping}. In
the initial period, Schwabe plotted spots without distinction of
umbrae and penumbrae. In the second period starting in 1831, Schwabe 
distinguished umbrae from penumbrae. In those cases, umbral sizes
were measured. We begin with the second period when areas are clearly
umbral and describe afterwards a work-around for the conversion of
sunspot sizes to umbral areas in the initial period until 1830.

\begin{figure}    
\includegraphics[width=0.49\textwidth]{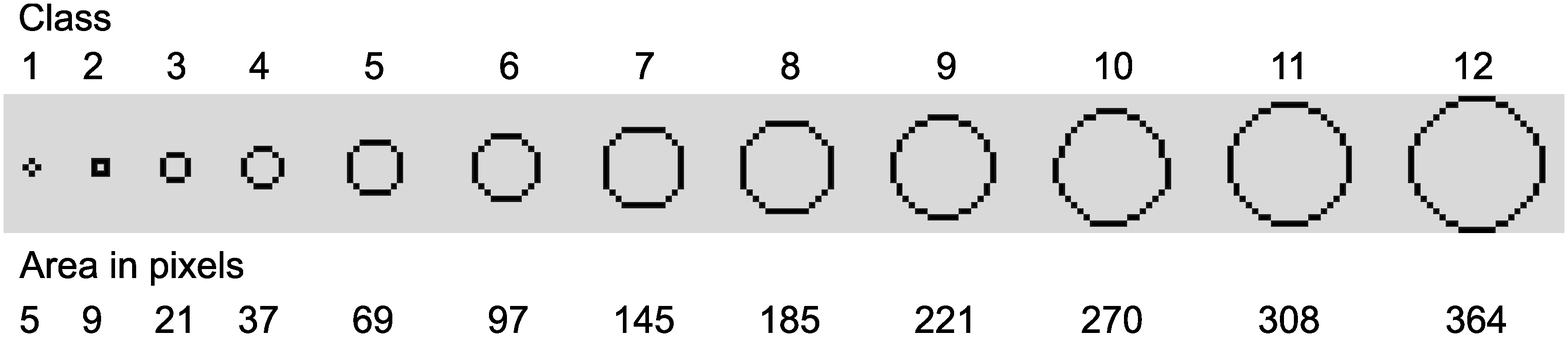}
\caption{Twelve cursor shapes (size classes) used for the size estimates of the sunspots
drawn by Schwabe. The bottom line gives the area in square pixels on the screen, including the black
border pixels.
\label{cursors}}
\end{figure}

\subsection{Indirect umbral areas for 1831--1867\label{indirectareas}}
One approach to obtaining physical areas of the sunspot
sizes is of a statistical nature. Sunspots were divided into 12~classes 
by size, as introduced by \citet{arlt_ea2013}. The measurements were 
actually carried out with twelve different circular cursor shapes having areas 
from 5~to 364~square pixels. Their shapes and screen 
areas are shown in Fig.~\ref{cursors}. The classes increase
monotonously in area, but were chosen relatively arbitrarily. 
Any more precise individual pixel-counts of umbral areas are unlikely 
to yield more accurate data, since the drawings are meant to be sketches 
of the sunspot distributions and sizes rather than exact copies. More 
details are given by \citet{arlt_ea2013}.

The division into 12~size classes is done both for Schwabe's 
dataset and for a modern data set (see below), whereby for the 
modern dataset the classes are chosen such that each class has 
the same relative number of sunspots as the corresponding 
class in Schwabe's data. In other words, the relative 
abundances of the twelve size classes of the Schwabe spots are 
compared and calibrated with 20th century data by building 
histograms of 12~artificial size classes constructed to contain 
the identical abundances. The average umbral area in such an
artificial class gives us the umbral area corresponding to a
Schwabe size class. Finally, a function for the area depending 
on the heliocentric angle of a spot from the centre of the solar 
disk is fitted for each size class. We will refer to this angle 
as `disk-centre distance' $\delta$ in the following.

The reference data sets used to obtain a statistical mapping of sunspot 
umbral areas are from photoheliographic data of the observatories of Debrecen, 
Mt. Wilson, Kodaikanal, and the MDI instrument of SOHO. As described 
by \citet{gyori1998}, an improved automatic analysis method was used 
for the Debrecen data starting in 1988. Before that, from 1974 to 1987, 
the areas were measured by video techniques \citep{Dezso_ea1987}. In the 
1974--1987 data, the area values of larger sunspots at disk-centre distances 
$\delta>60^\circ$ increase very rapidly with $\delta$. This effect is not seen in the 
area values measured from 1988 onward, so we have used the Debrecen data from 
1988--2013 only. The Mt. Wilson data were analysed by \citet{howard_ea1984}
and contain spot properties from 1917--1985. The Kodaikanal 
data covering 1906--1987 were obtained in almost the same way as the Mt. Wilson 
ones \citep{sivaraman_ea1993}. The MDI data for 1996--2010 were obtained 
using a telescope of 4 arc seconds resolution \citep{watson_ea2011} which is 
similar to or perhaps a bit worse than the resolving capabilities of 
Schwabe's setup. Note that the data from the Greenwich Photoheliographic 
Database were not used because it contains group area totals instead
of areas of individual spots.

Typical diameters of solar pores in white light are between 1000~km and 
4000~km \citep{keppens_martinez1996}. That converts to 
$0.8$--$13\cdot10^6$~km$^2$ or 0.26--4.1~millionths of the solar 
hemisphere (MSH).  We are using a lower limit of 1~MSH for 
the construction of the histogram as argued below. In the Debrecen
data, integer values of the projected area in millionths of the
solar disk are given ($0, 1, 2, \dots$), while the corrected areas
are given in MSH, also as integer values. Since the projection
correction increases the area, whereas the conversion to MSH reduces 
the value by half, the lower limits of 1 in both quantities are
therefore statistically compatible. We discarded all spots smaller 
than 1~MSH from the other data sets before the
analysis. Ideally, one would want to define a minimum spot size
plotted by Schwabe, but in reality, his drawing style was not
that straight-forward. Whenever he detected a small group on the
Sun, he indicated its location by small dots. In more complex
groups, however, he did not indicate all the small spots because 
of their abundance. There is apparently no clear lower limit
for the spot size. We therefore use a compromise at this point,
excluding the smallest pores, and start from 1~MSH which is also
the lower limit in the Debrecen data.

The relative abundances of the twelve cursor size classes, denoted 
by $i=1, 2, \dots, 12$, are determined for three different ranges 
of disk-centre distances, which were $\delta <30^\circ$, 
$30^\circ$--$60^\circ$, and $60^\circ$--$70^\circ$. These distance 
classes are numbered as $d=1,2,3$. The upper limit of $70^\circ$ 
is due to the fact that not all reference data sets contain spots 
beyond that distance. The four reference data sets (this number 
will be denoted by $N$ in the following) are now also divided into 
twelve classes fulfilling the same relative abundances as obtained 
for the Schwabe classes, again split into the three selected ranges 
of distances. The histograms are based on the umbral areas which are 
{\em corrected for foreshortening}. We therefore expect a mapping of
size classes with a fairly small dependence on $\delta$.

Then the area for each cursor size is calculated by the unweighted 
average of all spots
\begin{equation}
  \overline{A}_{id} = {\sum\limits_{n=1}^{N}\sum\limits_{j=1}^{S_{\!nid}} A_{njid}}
	\left/{\sum\limits_{n=1}^{N} S_{\!nid}}\right.,
  \label{areaaverage}
\end{equation}
where $\overline{A}_{id}$ is the area for a cursor of $i$-th size class 
in the $d$-th disk-centre distance class, ${A_{njid}}$ is the umbral 
area (corrected for projection by $\cos^{-1}\delta$) of the $j$-th 
spot in the $n$-th data source, $i$-th size class and $d$-th distance 
class, and $S_{\!nid}$ is the total number of spots present in $n$-th 
data source, $i$-th size class in $d$-th distance class. Note that these
averages consist of different histogram bins for different $n$. For
example, the equivalent class-5 bin in the Debrecen data will have other area 
limits than the equivalent class-5 bin of the MDI data. The averaging helps smooth
possible systematic over- or underestimations of areas in the 20th-century
data sources. The averages $\overline{A}_{id}$ are not immediately
areas corrected for foreshortening, since the histogram classes are 
constructed using Schwabe's raw sunspot sizes. We will capture any possibly remaining
disk-centre distance dependence in Sect.~\ref{distance}, where functions
through the three distance classes for each $i$ will be derived, i.e.
12~functions for the 12~size classes.

\subsection{Indirect umbral areas for 1825--1830}
The sunspots in the early full-disk drawings from 1825 to 1830 were not
drawn at a good resolution. Instead, Schwabe plotted high-resolution 
magnifications at unknown scale beside the full-disk drawings. The 
magnifications show that nearby spots were combined in the full-disk 
drawings.  The spots in these drawings are therefore often `blobs' made out 
of very close spots and including the penumbrae between those spots. 
Hence, the pencil dots used to measure the sizes of those spots do not 
represent their umbral area. To estimate the area for these spots, 
we need to compare the size statistics with grouped spots including penumbrae. 
To do that, we combined the spots inside a single penumbra in a modern data 
set and used these for the statistical estimation of the area. 

Among the data used here, the Debrecen data is the only source which 
contains umbral and penumbral areas broken down into individual spots. Recently,
\citet{tlatov_ea2014} published detailed measurements of the Kislovodsk Mountain
Astronomical Station. Since those only cover the somewhat peculiar cycle~24, we
prefer not to use them for the area calibration of Schwabe's drawings. The 
conversion of the 1825--1830 data is therefore based on the Debrecen data 
(1988--2013) as a mixture of different cycles. From that source, we prepared a data set in which 
all the umbrae inside a continuous penumbra are added and considered 
as a single spot of which we store the whole-spot area and the total 
umbral area in each spot. Now we divide the {\em whole-spot areas\/} into 
12~classes with the same relative abundances as we obtained from Schwabe's 
1825--1830 data, but use the corresponding {\em umbral areas\/} for an average 
according to Eq.~(\ref{areaaverage}) for each size class in each distance 
class. The results for the distance classes are again combined into a 
function of the disk-centre distance in Sect.~\ref{distance}.

There are still differences between these combined spots from the Debrecen data and 
the pre-1831 Schwabe spots. (1) A penumbra with a single umbra would look similar to
a penumbra with two umbrae in Schwabe's drawings. The penumbral area between the two
umbrae in the latter case, however, will lead to different total umbral areas when derived 
from the Debrecen data. (2) It is not always the entire penumbra which Schwabe plotted
in a `blob'. (3) All the umbrae inside a penumbra are added in the Debrecen data, even
for very extended penumbrae. Schwabe, however, did not club together all the umbrae 
inside a connected penumbra when it was very large, but plotted `sub-penumbrae' in that 
case.

There are 38~spots apparently drawn without the inclusion of a penumbra 
which were found from the visual comparison of disk drawings with magnifications. 
These spots were not used in the procedure described here; their areas were 
calculated using the method discussed in Sect.~\ref{indirectareas}.

\begin{figure}    
    \includegraphics[width=0.49\textwidth]{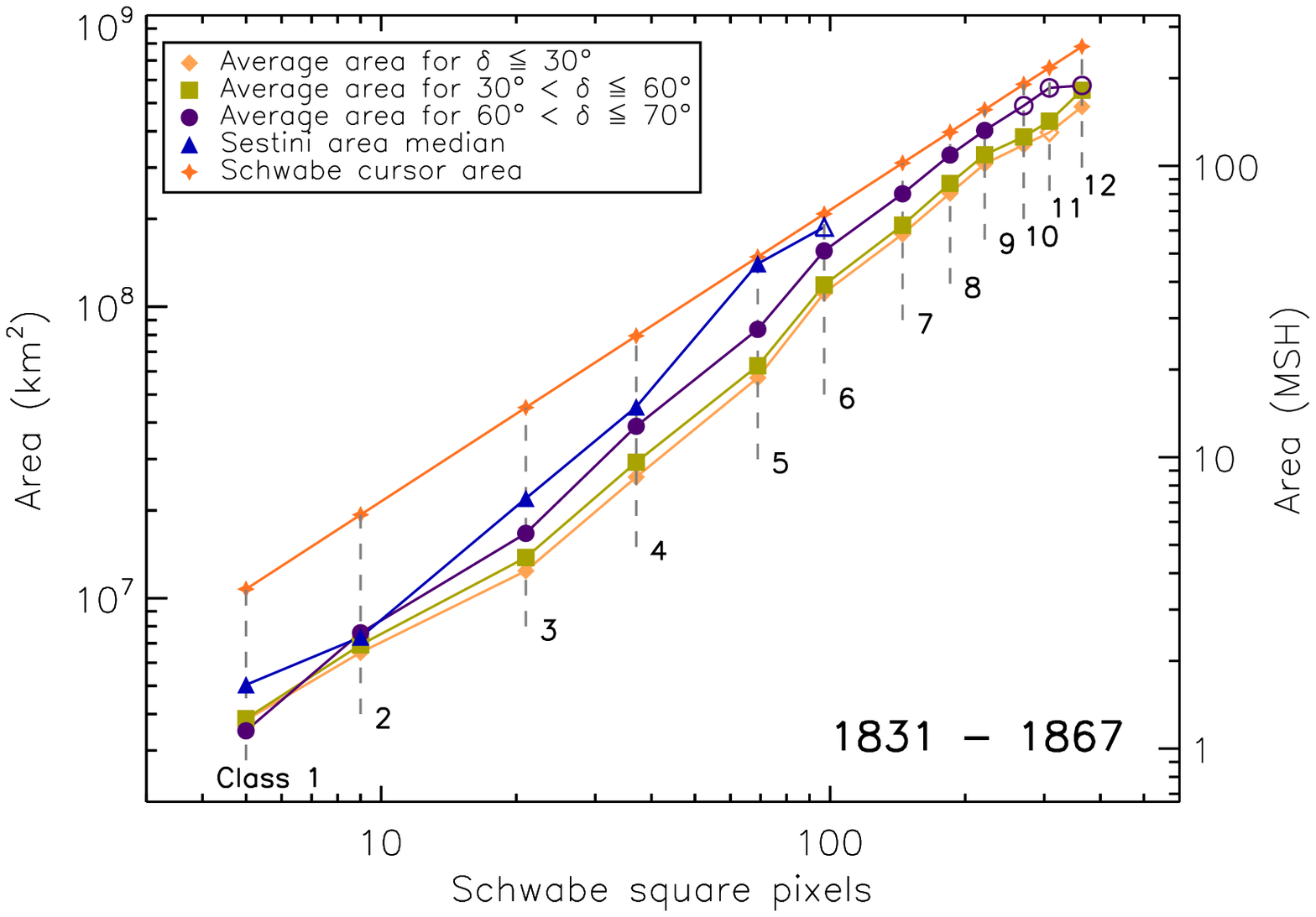}
    \includegraphics[width=0.49\textwidth]{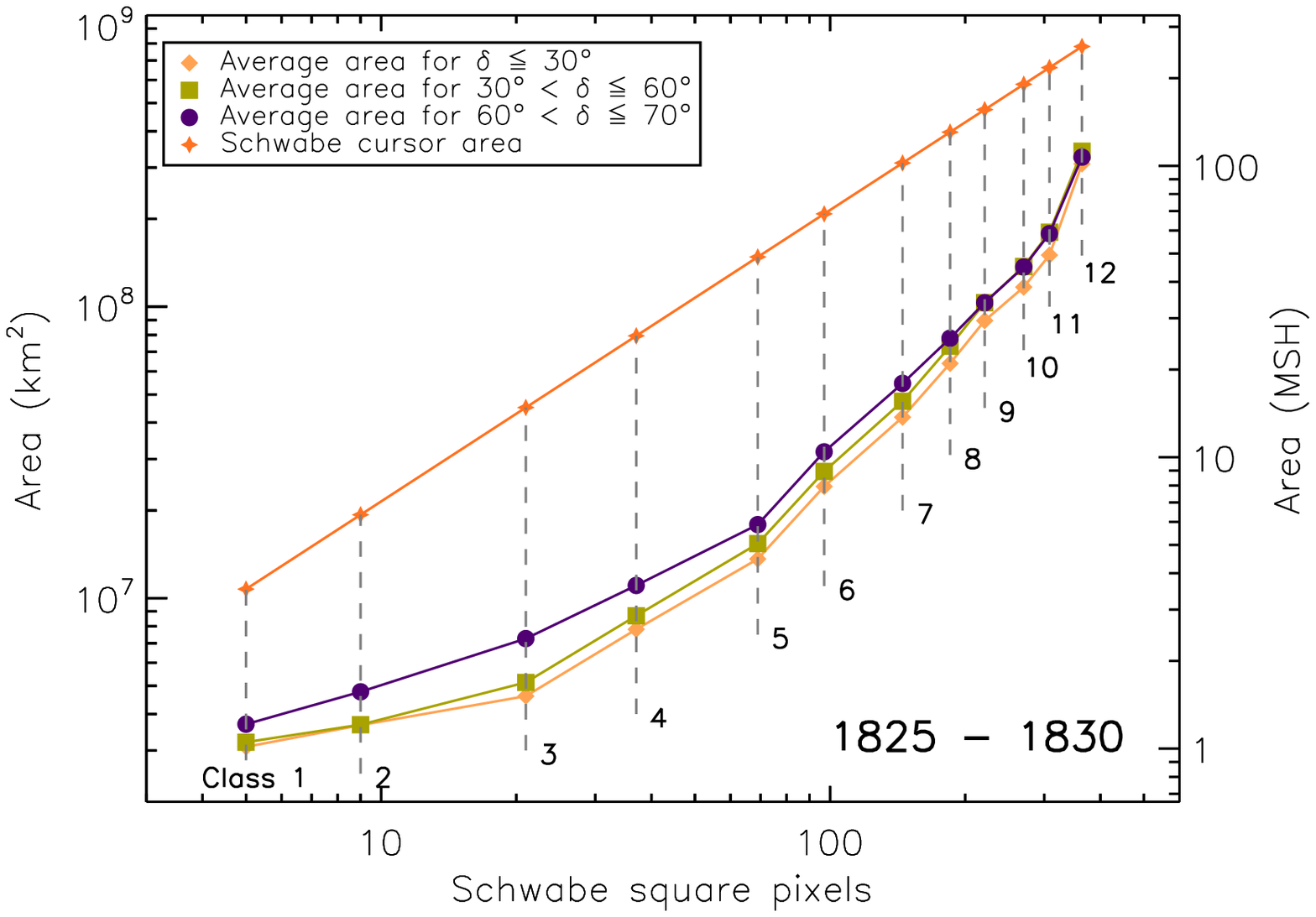}   
\caption{Average areas in km$^2$ (left ordinate) and MSH (right ordinate) 
for each of the twelve size classes (abscissa), and for three different 
centre-distance ranges: $<30^\circ$ (diamonds), $30^\circ$--$60^\circ$
(squares), and $60^\circ$--$70^\circ$ (circles). The areas corresponding to
the simple pixel areas of the size classes without any
calibration are indicated by plus signs. The direct calibration of size 
classes by observations from Sestini in 1850 are shown as triangles. {\em Top:\/} 
average areas for the data from 1831--1867. {\em Bottom:\/} average areas for 
the data from 1825--1830. An open symbol means that the number of 
spots used in the calculation of the average area is less than twenty.
Note that the lower curves in the lower panel {\em do not imply\/} smaller
areas, but they rather mean that any {\em spot of a given true area\/} were drawn as
a spot of larger class in 1825--1830 than in 1831--1867.}
\label{area_versus_size}
\end{figure}

\subsection{Umbral areas from parallel observations}
For a short period from 1850 Sep~19 to Nov~4, large-scale drawings are 
also available from Sestini who observed from Washington, D.C., 
comprising a total of 42~full-disk graphs \citep{sestini53}. We compiled 
spots seen by both Schwabe and Sestini and measured the umbral 
areas of the corresponding spots in Sestini's drawings. There 
were not enough observations to cover the whole range of size 
classes -- we could only cover classes 1--6. We will come 
back to the results in the following Section. One has
to note though that the time difference between the observations
from Dessau and the ones from Washington implies changes in the
evolution of the
spots either leading to wrong areas or to wrong spot associations
between the two observers.

\subsection{Final mapping of sunspot sizes\label{distance}}
Figure~\ref{area_versus_size} shows the mappings of Schwabe
size classes to physical areas in km$^2$ and MSH for three 
different ranges of disk-centre distances for the statistical 
conversion as well as a mapping for the calibration 
with concurrent high-resolution observations. One 
immediate result is that the direct conversion of
the pencil spots in Schwabe's drawings into sunspot areas would lead to
overestimated umbral areas in most size classes. One
can also see that the areas corresponding to the pixel
areas do not form a linear function or power law. The areas from the comparison
with the sunspots observed by Sestini in 1850 are in good agreement
with the indirect mapping based on the size distribution. This shows that the direct
conversion of pixel areas into sunspot areas is not a good choice. The
only exception is class~5, but it contains only 21~measurements 
and may be a poor estimate (as is the one for class~6).

The lower panel of Fig.~\ref{area_versus_size} appears to 
show much smaller areas. This is not true, however. The 
graph actually shows that a 5-MSH spot which was typically 
drawn as a class-3 dot in 1831--1867, was represented by a 
class-5 dot before that period, since it encompassed a 
considerable fraction of the penumbra at that time. The areas
of class~3 in the top panel cannot be compared with the areas 
of class~3 in the lower panel. Sunspot sizes corresponding 
to large size classes were much more often used by Schwabe 
in the period of 1825--1830 than afterwards (note that there are 
no longer open symbols with fewer than twenty spots in the lower panel). 

The dependence of final areas on the disk-centre distance
is described by functions of the form
\begin{equation}
  {\cal A}_i(\delta) = a_i + \left(b_i/ \cos\delta\right),
  \label{mapping}
\end{equation}
where $a_i$ and $b_i$ are coefficients for the $i$-th size class
and $\delta$ is the distance of the spot from the centre of the solar 
disk. In the end, there are 12~functions for the period 1825--1830 
and another 12~functions for the period 1831--1867. They deliver a
mapping of the size classes into physical areas.

When computing the areas for the final sunspot data base, the area 
is not calculated if a spot distance is greater than $85^\circ$ 
from the disk centre, since the area values become very uncertain. 
All spots with $\delta\leq70\degr$ are reliable in the sense that they 
are covered by the statistics leading to the mapping. All spots with 
$70\degr<\delta\leq85\degr$ are uncertain because they rely on an 
extrapolation of the mapping, while all spots with $\delta>85\degr$ are 
highly uncertain and are therefore excluded from the data base. 
The smallest sunspot area occurring in our data after applying 
Eq.~(\ref{mapping}) is 1~MSH which is consistent with the initial 
lower limit for spots in the reference data sets.

\begin{figure}        
    \includegraphics[width=0.49\textwidth]{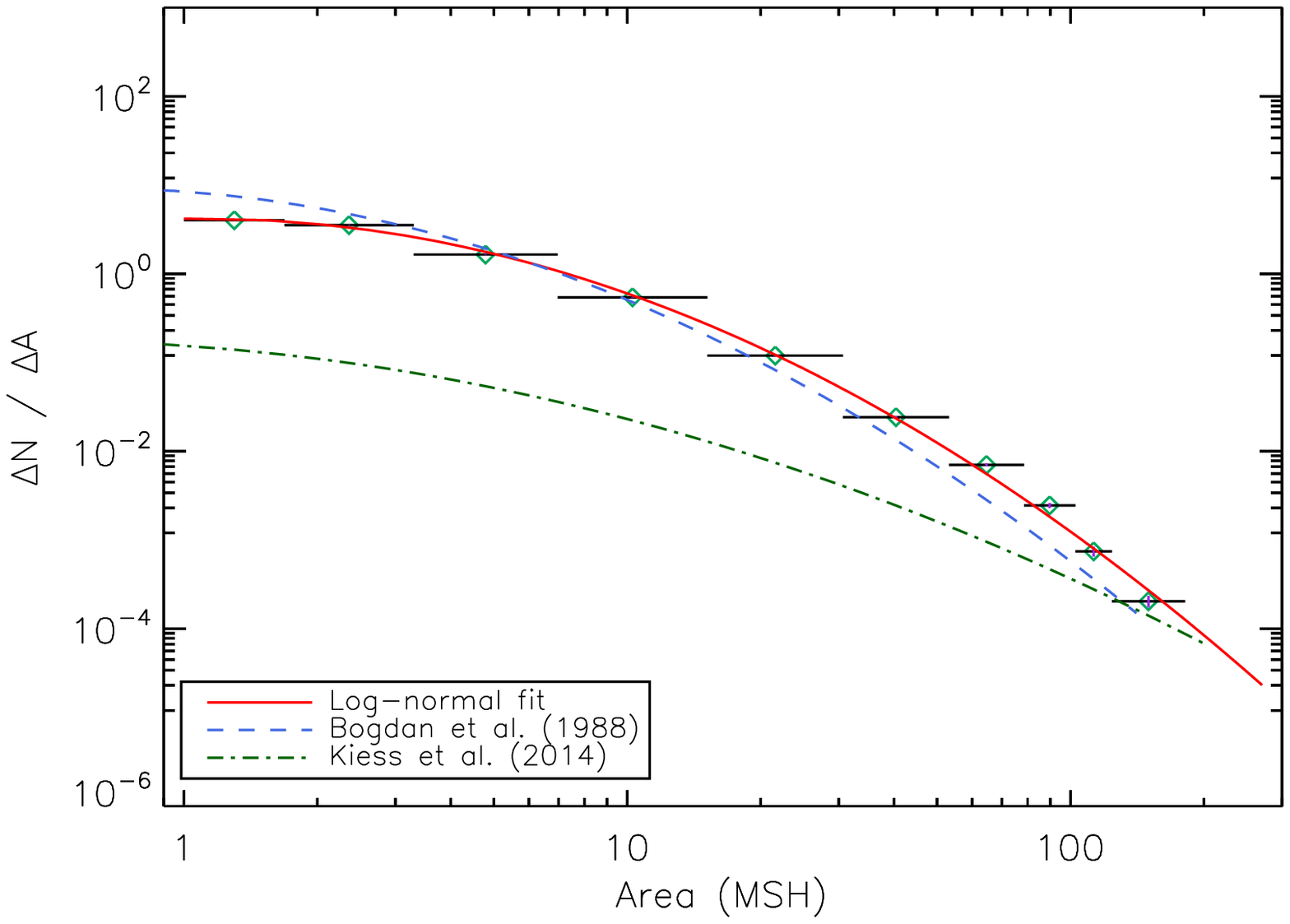}    
    \includegraphics[width=0.49\textwidth]{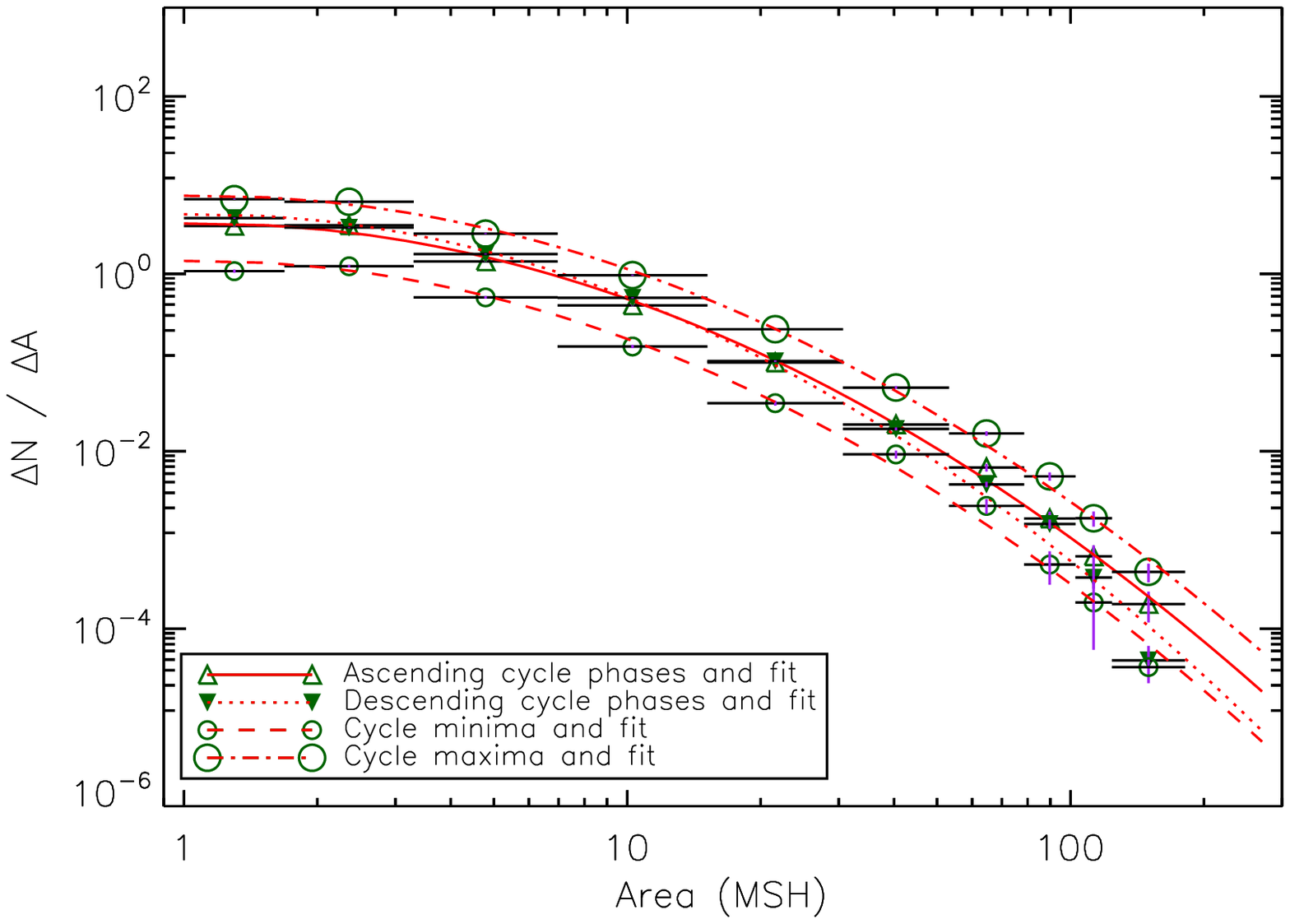} 
\caption{The distribution of estimated umbral areas of sunspots. 
{\em Top:} The area distribution for 1831--1867. The solid curve shows
the log-normal fit. The dashed curve and dash dot curve are the fit curves
form \citet{bogdan_ea1988} and \citet{kiess_ea2014}, respectively. The 
horizontal bars show the widths of the bins. The error margins on the 
distribution values are all smaller than the symbols. {\em Bottom:} 
The area distribution of spots for ascending phases, descending phases, 
cycle minima and maxima of all cycles within 1831--1867. The lines show the log-normal fits 
for the corresponding area distributions.}
\label{area_dist}
\end{figure}

\subsection{Distribution of sunspot area}

The data base of Schwabe's observations contains a total of 
135\,921~entries comprising 134\,386~spots with size estimates 
(each line corresponds to an individual umbra) as well as
1535~spotless days (each line corresponds to a day with zero spot size).
Whenever we use the term `spots' we refer to individual umbrae as far as 
Schwabe resolved them. Positions are not available for 849~umbrae
because the orientation of the drawing could not be identified. No
physical areas are available because $\delta$ is missing.
There are another 41~umbrae beyond $85^\circ$ from the centre of the disk 
for which areas are not calculated because of too large uncertainties. 
The area values are therefore available for 133\,496~umbrae. To study the 
distribution of areas, we consider the spots within $\pm 50^\circ$ central 
meridian distance (CMD) and within $\pm 45^\circ$ latitude. The same latitude 
limit was chosen by \citet{bogdan_ea1988}. The central meridian limit, 
however, was $\pm 7\fdg5$ in \citet{bogdan_ea1988} in order to avoid 
duplicate counts of groups (a group will typically appear only once in 
a $15\degr$ window because of the solar rotation). Since drawings by 
Schwabe are not available on all days, we need to widen this window 
reasonably and found $\pm 50^\circ$ a good compromise between not missing 
too many groups due to observing gaps on the one hand and the 
contamination by duplicate consideration of groups on the other hand. 
The latter will actually not affect the distribution significantly, 
since \citet{baumann_solanki2005} and \citet{kiess_ea2014} 
did not see any drastic changes between counting 
umbrae only once and counting them on every day they were visible. The lowest area 
considered for the distribution is our lowest area, 1~MSH. The above 
criteria reduce the data to a sample of 104\,217~spots in total, and 
96\,984~spots from 1831 to 1867.  
 
The umbral area spectrum was obtained as described by \citet{bogdan_ea1988}
with the exception that they used a lower umbral area limit of 1.5~MSH.
Since the distribution is differential, the different lower limit should
not affect the shape of the curve.
The bins for small areas were selected such that each bin encompasses 
approximately one size class up to class~9, whereas one additional bin was defined
such that it contains the spots from class~10 and about half the spots in class~11. All spots even larger than
that ($\geq 185$~MSH) were outside the above mentioned CMD window. Dividing the area range 
into twenty bins as in \citet{bogdan_ea1988} would have caused a strong scatter
since the Schwabe areas accumulate near twelve typical area values, because the
dependences on the disk-centre distance described by (\ref{mapping}) are all
small.

The area distribution of the Schwabe spots also resembles a log-normal 
distribution and looks similar to the curve by \citet{bogdan_ea1988}.
The parameters for such a distribution over the area $A$ are obtained 
through a fit by the function
\begin{equation}
  \ln\left(\frac{\mathrm{d}N}{\mathrm{d}A}\right) = -\frac{\left(\ln~A  - \ln~\langle A \rangle\right)^2}
	{2~\ln~\sigma_{\!A}} + \ln\left(\frac{\mathrm{d}N}{\mathrm{d}A}\right)_{\rm max},
\end{equation}
where $(\mathrm{d}N/\mathrm{d}A)_{\rm max}$ is the maximum of the area distribution function, 
$\langle A \rangle$ is the mean, and $\sigma_{\!A}$ is the geometric standard 
deviation. Table~\ref{fitparameters} shows the log-normal fit parameters 
obtained with a Levenberg-Marquardt least-squares method. The 
cycle minima and maxima were taken from the ``Average'' column of the
cycle timings by \citet{hathaway2010}. Minima periods and maxima periods
are defined as $\pm1$~yr around the minima/maxima, while the ascending
and descending phases are the remaining periods.

\begin{table}
\caption{Log-normal fit parameters for the Schwabe data and various 
subsets of them.}
\begin{tabular}{lrccc}
\hline
\hline
Data & Umbrae & $\langle A \rangle$ & $\sigma_{\!A}$ & $\left(\frac{\mathrm{d}N}{\mathrm{d}A}\right)_{\rm max}$ \\
     &        &  [MSH]               &  [MSH]        &    [MSH$^{-1}$] \\
\hline
All data (1825--1867)       & 104\,217 & 1.05 & 3.8 & 3.8 \\
1825--1830                  & 7\,233   & 0.58 & 9.9 & 1.7 \\ 
1831--1867                  & 96\,984  & 1.10 & 3.5 & 4.2 \\
\hline
Cycle 7                     & 9\,448   & 1.09 & 5.3 & 1.3 \\
Cycle 8                     & 22\,382  & 1.08 & 3.8 & 4.0 \\
Cycle 9                     & 36\,862  & 1.08 & 3.0 & 5.3 \\
Cycle 10                    & 35\,181  & 1.10 & 3.4 & 4.9 \\          
\hline
Ascending phases            & 13\,613  & 1.09 & 3.5 & 3.7 \\
Descending phases           & 46\,763  & 1.10 & 3.1 & 4.7 \\
Cycle minima                & 3\,507   & 1.08 & 3.4 & 1.4 \\
Cycle maxima                & 31131  & 1.10 & 3.6 & 7.6 \\
\hline
\label{fitparameters}
\end{tabular}
\end{table}

The top panel of Fig.~\ref{area_dist} shows the resulting total area 
distribution of umbrae for 1831--1867. The errors on the ordinate 
values were estimated by $(\Delta N/\Delta A)/\sqrt{\Delta N}$, where 
$\Delta N/\Delta A$ is the discrete area distribution value and 
$\Delta N$ is the number of spots in each bin. The errors are all smaller 
than the symbols. The lower curve from Fig.~1 in \citet{bogdan_ea1988},
which is the fit to the full range of umbral areas of 1.5--141~MSH,
and the curve from \citet{kiess_ea2014} are also plotted in 
Fig.~\ref{area_dist} for comparison. While the data of both analyses also
influence our size calibration of Schwabe's sunspots, the distribution
only agrees with the one by \citet{bogdan_ea1988} based on Mt. Wilson
data. Interestingly, also the area distributions obtained from group
umbral areas \citep{baumann_solanki2005} agree fairly well with our
results. The peak position $\langle A\rangle$ of their log-normal distribution from the
Greenwich group data is ten times larger than ours from individual 
spot data, in good agreement with the fact that sunspot groups consist 
of roughly ten spots on average.

The bottom panel of Fig.~\ref{area_dist} shows four individual area 
distributions for the ascending and descending phases of the cycles 
as well as for the cycle minima and maxima. They exhibit 
nearly the same distribution as the whole area distribution. The descending 
phases of all cycles have larger $(\mathrm{d}N/\mathrm{d}A)_{\rm max}$ than the ascending phases,
however. The fairly small variation of $\sigma_{\!A}$ is remarkable and 
confirms the corresponding findings by \citet{bogdan_ea1988} and
\citet{schad_penn2010}.

The umbral area distribution was calculated for the data from 1825--1830 and 
from 1831--1867 separately, and the fit parameters are given in 
Table~\ref{fitparameters}. The distribution for the earlier period
is much wider, with $\sigma_{\!A}$ being similar to the one by \citet{kiess_ea2014}
who derived $\sigma_{\!A}=10.7$ ($\sigma=1.54$ in their work).

\section{Group definitions\label{groups}}
Tilt-angles of groups will be sensitive to the actual association 
of spots into groups. Schwabe's original drawings contain group 
names which he started from number one every new year. His 
perception of a group was often too broad. A fair number of 
sunspot clusters actually contain two or more groups. 
This new definition of the groups was made by manual inspection 
of the drawings. We also used the evolutionary information of the 
clusters and sub-clusters provided by the images of adjacent days. 
Very often, a small apparently new bipolar group emerged near an 
existing one and showed its individual evolution through the Waldmeier 
(or Zurich) types. Schwabe included them in the group number of the 
existing group, while we defined a new group in many such cases.
In other cases, when splitting of the polarities was not obvious
and the parts of the group were all in the same evolutionary phase,
we kept Schwabe's definition, despite leading to somewhat large
groups of $30\degr$ extent or more. Any splitting, however, would have
been very arbitrary and would add noise rather than new information 
to the tilt angle data base. A total of 56~groups with extents
$\geq30\degr$ size remained.

\begin{table*}
\caption{Statistics of coverage, spot areas, groups, and tilt angles $\alpha$ for Schwabe's
observations in 1825--1867. CMD is the central meridian distance of the
area weighted centre of a group.\label{statistics}}
\begin{tabular}{lrrrrrrr}
\hline
\hline
Period     & Days with     & Gaps longer & Gross groups& Unique groups & Groups    & Group tilts with     &Group tilts with      \\
           & drawings      & than 5 days\tablefootmark{a} & with areas  & with areas    & with tilt & $|{\rm CMD}|<60\degr$& $|{\rm CMD}|<60\degr$\\
           &               &             &             &               &           &                      & and $\alpha > 3\degr$\\
\hline
1825--1830 & 1187 (63.0\%) &    32\phantom{$^2$}& 4\,401 &  945       &  2\,452     &  2\,159                & 1\,745\\
1831--1867 & 8808 (65.2\%) &   149\phantom{$^2$}&27\,116 & 5903       & 20\,689     & 17\,365                &13\,803\\
\hline
Total      & 9995 (64.9\%) &   182\tablefootmark{b}&          31\,517 & 6848       & 23\,141     & 19\,524                &15\,548\\
\hline\\[-1ex]
\end{tabular}
\tablefoot{
\tablefoottext{a}{The gaps are derived only from the
days for which we obtained data; if we include the unused drawings, the number of
gaps longer than five days is a bit smaller than the number given in this column.}
\tablefoottext{b}{One group is missing in the 1825--1830 number because a gap straddles 1830 and 1831.}
}
\end{table*}

\begin{figure}  
\includegraphics[width=0.49\textwidth]{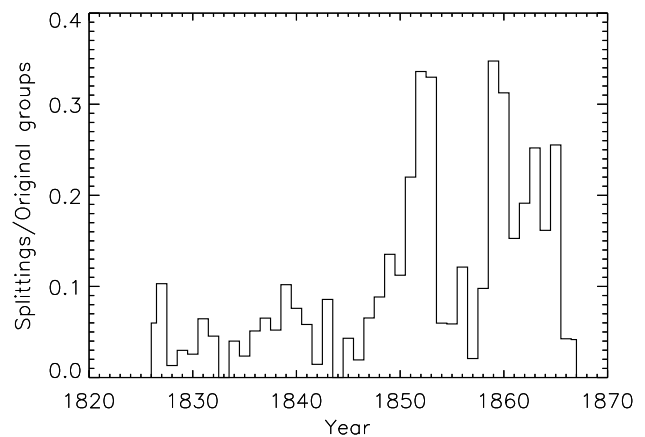}
\caption{The number of new groups obtained by splitting Schwabe's 
original groups, normalised to the number of groups before splitting 
(i.e. Schwabe's original groups). The criteria for splitting a group 
are described in the main text.}
\label{split_hist}
\end{figure}

\begin{figure} 
\includegraphics[width=0.49\textwidth]{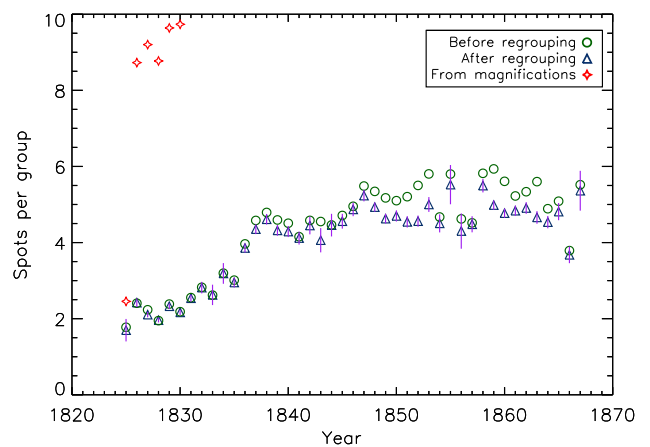}
\caption{Annual averages of the number of spots per group calculated 
before regrouping of sunspot groups (circles), after regrouping (triangles), 
and manually counted number of spots from the magnification drawings of 
sunspot groups (diamonds). Uncertainties are only given for the values 
after regrouping and are obtained from the relative Poissonian
error $1/\sqrt{n_{\rm counts}}$, where $n_{\rm counts}$ are the number
of all instances of all groups in a given year (groups count several times
with different numbers of spots).}
\label{spt_grp_year}
\end{figure}

\begin{figure} 
\includegraphics[width=0.49\textwidth]{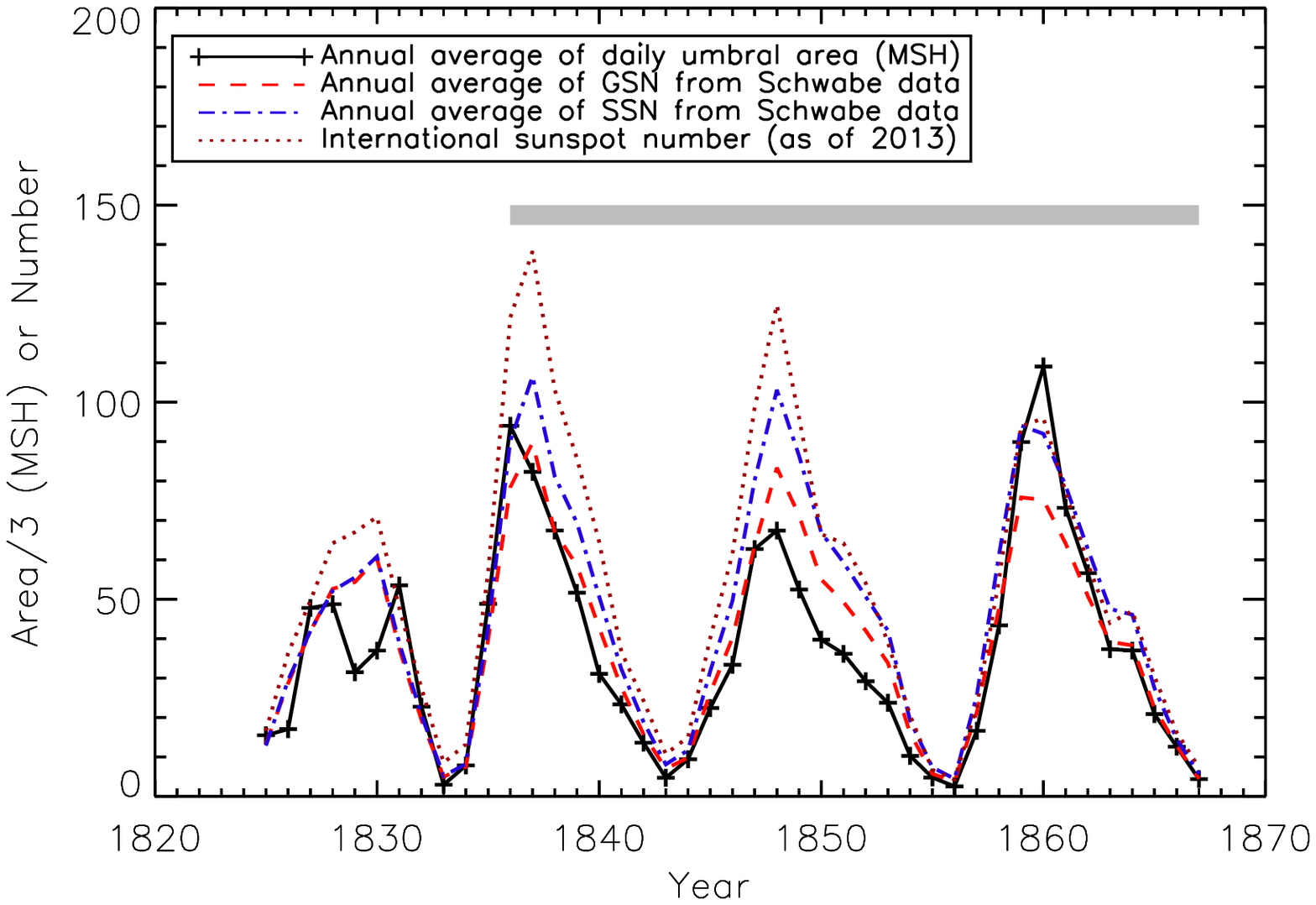}
\includegraphics[width=0.49\textwidth]{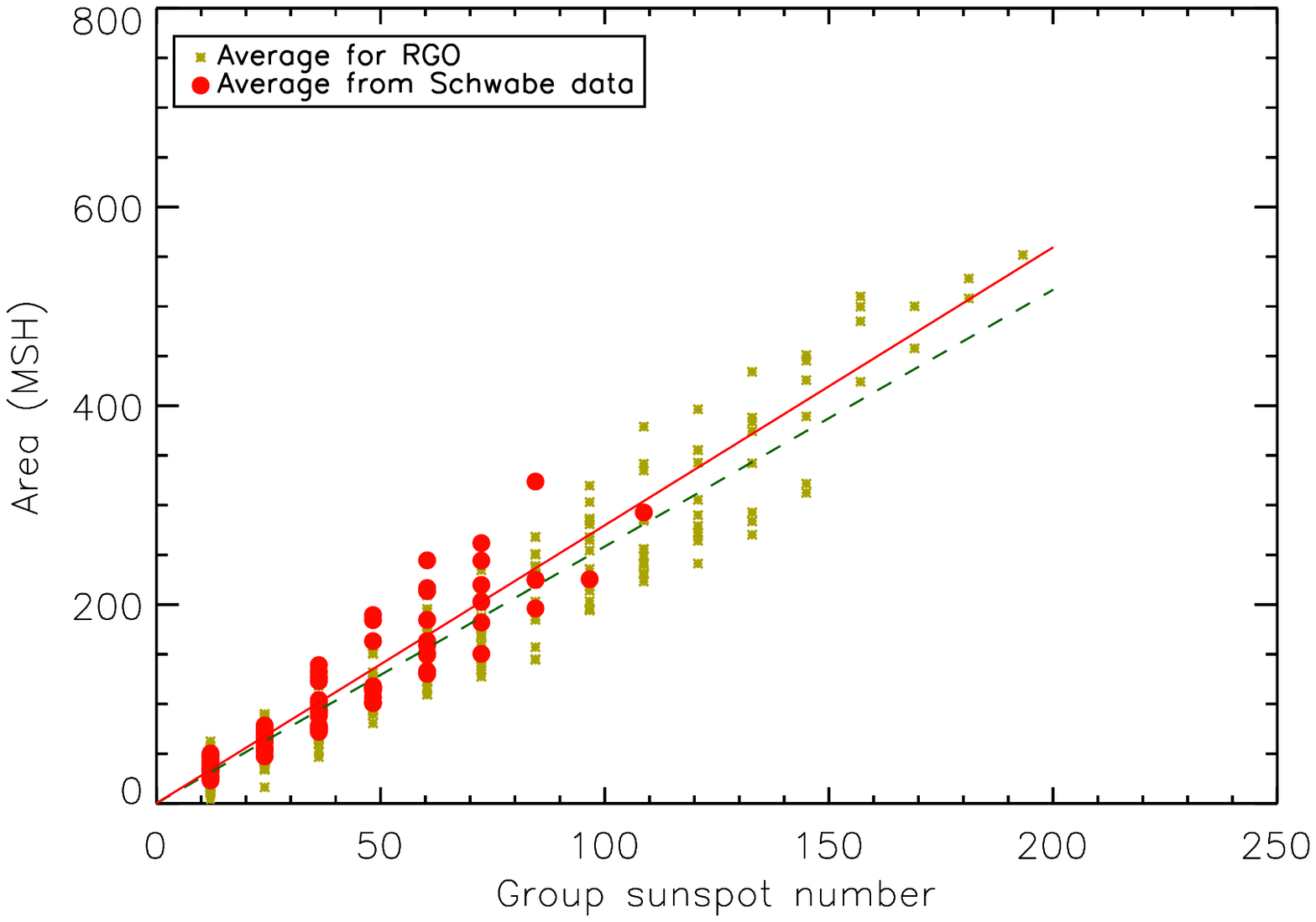}
\caption{{\em Top:\/} Annual averages of total-disk umbral area of sunspots 
in MSH and divided by three (solid line), the group sunspot number (GSN) derived from our 
groupings (dashed line), the sunspot number (SSN; Wolf number)
derived from our groupings and the actual number of spots in the
full-disk drawings (dash-dotted line), and the official International
Sunspot Number (dotted line) of before the revision suggested by 
\citet{clette_ea2014} which was not yet available. The gray horizontal
bar indicates the period in which the number of spots per group was
fairly constant (cf.\ Fig.~\ref{spt_grp_year}). {\em Bottom:\/} 
Daily total umbral areas versus GSN (in case of Schwabe, the GSN
is based on our grouping in Schwabe's drawings, corresponding to the 
dashed line in the upper panel), averaged over 100~values in 
each group sunspot number bin for the Schwabe data (circles) and the Greenwich 
data (asterisks). The solid and dashed lines are linear fits through 
the Schwabe and Greenwich data, respectively, being enforced to go through
the origin.}
\label{yearareaavg_gsn}
\end{figure}

Figure~\ref{split_hist} 
shows the ratio of the number of newly formed groups, i.e. those 
obtained by splitting Schwabe's original groups, to the original
number of groups. For instance, if there were a total of 10~groups 
which were all split into two, resulting in 10~new groups, the 
ratio would be one (100\% splitting). The large fraction of splittings seen after 
1850 is partly due to the presence of very many closely located groups,
but chief\/ly due to a wider definition of groups Schwabe adopted 
during those cycles. In nine cases, Schwabe assigned one group designation
each to two spots, while they apparently form one bipolar
group. We combined those cases to one group name.

The splittings are marked by modified group names in the above mentioned data 
file. The two groups 116 and 117 in Fig.~\ref{penumbrae} will now appear as
four groups, 116-0, 116-1, 117-0, and 117-1 in the catalogue. Combined groups
are marked with plus signs in the new group name, e.g. 39+40.

To demonstrate the impact of the regrouping (splitting as well as combining), 
Fig.~\ref{spt_grp_year} shows the annual averages of the number of spots per 
group, derived from Schwabe's group definitions and derived after regrouping
by visual inspection. Additionally, the sunspot group magnifications drawn by
Schwabe (see Fig.~\ref{clumping}) were used to compute a third set of average spot 
numbers per group.  Since the magnifications are not biased towards exceptionally large groups 
until 1830, we give only the averages for 1826--1830. These spot numbers per group 
obtained from detailed drawings of individual groups match modern values very well 
\citep{tlatov2013,clette_ea2014}. After 1830, only selected, big groups were
magnified, so the values of spots per group are biased.

In the averages derived from the full-disk drawings, there is a significant 
increase in the number of spots per group from 1830 to 1836. This increase 
was not flattened after the regrouping of sunspot groups. It may be partly 
due to initially smaller true numbers of spots per group and partly due to the early, coarser
drawing style of Schwabe. The most notable jump from 1835 to 1836 does not
coincide, however, with the change in drawing style in 1830/1831.
An increase in the number of spots per group can also be seen after the other minima, namely
in 1843--1847, 1856--1858, and 1866--1867. Therefore, the change in drawing style
and the recovery from the activity minimum in 1833 are probably
superimposed effects.

By the same token, we may spot small peaks coinciding with the solar 
cycle maxima~8, 9, and~10 in Fig.~\ref{spt_grp_year}. 
This effect has also been observed -- even
more drastically -- by \citet{clette_ea2014} in 20th-century data, 
and it may actually be a mixture of a real effect and observational 
bias (basically because on a crowded Sun, the splitting of groups 
is difficult).

The upper panel of Fig.~\ref{yearareaavg_gsn} shows the annual averages 
of umbral areas of sunspots. They are compared to the yearly averages of 
the group sunspot number (GSN) according to our own group number information 
and the (Wolf or Zurich) sunspot number (SSN), both derived from Schwabe's 
observations, as well as to the International Sunspot Number (ISN).\footnote{Note that the
GSN includes a scaling of 12.08 derived from the comparison of the
ISN with the groups found in the Greenwich Photoheliographic Database
\citep{hoyt_schatten1998}. Since the ISN was scaled down to match
Wolf's observations, who recorded about 60\% of the sunspots that would
be reported today, the SSN from Schwabe's data can actually lie above
the GSN. This is the case when Schwabe's drawings are bit more detailed
than Wolf's reports.} Good
agreement is found between umbral areas and Schwabe's group sunspot
number for cycle~8, while the areas of cycles~7 and 9 fall below the
(rescaled) sunspot numbers and cycle~10 has larger areas than the
sunspot numbers indicate. Since Schwabe's observing method,
telescope and drawings are very constant after 1835, the difference 
between the ISN and the Schwabe record may be due to calibration issues 
of the ISN before 1849 \citep{leussu_ea2013}.

In an attempt to assess the correlation of the umbral areas with the 
group sunspot number, we plot averages of 100~daily all-disk umbral 
areas versus the corresponding group sunspot number of the same days
in the lower panel of Fig.~\ref{yearareaavg_gsn}. The same was done for
the Greenwich photoheliographic database which contains umbral areas 
until 1976. The graph is similar to the one by \citet{balmaceda_ea2009} 
who used the total spot areas instead. There is an intrinsic scatter 
in the correlation due to a certain randomness if both the sunspot 
number and sunspot areas are related to an internal magnetic field, rather 
than to each other. The uncertainty from the randomness has been reduced 
to about 10\% by averaging over 100~days. Although the scatter in total 
umbral areas is higher than that due to randomness (as seen in the 
bottom panel of Fig.~\ref{yearareaavg_gsn}), it is comparable to the 
results from the Greenwich data, except for a slight tendency to 
larger areas (lines of linear fits through the origin were added for 
clarity in Fig.~\ref{yearareaavg_gsn}). We therefore conclude that 
the areas inferred from the Schwabe data are compatible with the 
Greenwich data, which did not enter our calibration at any step. The 
Schwabe areas do show, however, cycle-to-cycle variations in the
strength of the correlation with the sunspot number indices.

The final total numbers of groups as well as the numbers of groups
that will have tilt angles (see Sect.~\ref{tilts}) are given in
Table~\ref{statistics}.
  
\section{Tilt-angles of groups}\label{tilts}
\subsection{Determination of tilt and separation}
The tilt angle of a given sunspot group is calculated in a plane
tangential to the solar surface in an estimated mid-point of that 
particular group to avoid problems with the curvilinear heliographic 
coordinates. The mid-point of the group (hereafter box-centre as opposed to the
area-weighted centre of gravity of the group) is obtained 
using the easternmost and westernmost spots as well as the northernmost 
and southernmost spots of a given group. The longitude and latitude of the 
box-centre is set to be the contact point of the tangential plane with 
the solar surface. The Cartesian coordinates in this plane
are $x_i$ and $y_i$ for the $i$-th spot and $x_{\rm g}$ and $y_{\rm g}$
for the box-centre and are normalised with respect to the solar diameter.

The algorithm then checks for the number of spots in a group. If it 
is equal to two, it proceeds to calculate directly the tilt angle 
and polarity separation. If it is more than two spots, we have to 
assess the most probable configuration of which spots belong to which 
polarity, since magnetic information is not available. We look for the 
most probable division of the group into two clusters by finding the 
least positional variance within the individual clusters.  For that, we let 
a division line, running through the box-centre, rotate from $\theta=0$ to 
$\theta=180\degr$ and obtain -- for each angle -- a cluster of `leading' spots 
and a cluster of `following' spots. This is achieved by using a vector perpendicular 
to the division line, $\vec{D} = \left(\cos~\theta,\sin~\theta\right)$.
The sign of the inner product of this vector with the spot vector, 
$\vec{S}_i = (x_i - x_g,y_i - y_g)$ defines the cluster (`polarity') membership 
of the $i$-th spot. The sum of the two variances of the spots' 
coordinates on either side of the division line is calculated. We denote the angle 
at which the least positional variance is achieved by $\theta_{\rm opt}$ 
and adopt it as the most probable division of the sunspot group into 
polarities. The area-weighted centres of the polarities found are then
calculated. The eastern and western parts correspond to the following and leading 
polarities, respectively, and their coordinate pairs are denoted by
$(x_{\rm F},y_{\rm F})$ and $(x_{\rm L},y_{\rm L})$. The coordinate 
pairs convert to heliographic coordinates $(\phi_{\rm F},\lambda_{\rm F})$
and $(\phi_{\rm L},\lambda_{\rm L})$, respectively.

The tilt angle $\alpha$ is computed by
\begin{equation}
\tan \alpha=\left\{\begin{array}{l}
\left(y_{\rm F}-y_{\rm L}\right)/\left(x_{\rm L}-x_{\rm F}\right)\qquad\mbox{if $\lambda_{\rm g}\geq0$}\\
\left(y_{\rm L}-y_{\rm F}\right)/\left(x_{\rm L}-x_{\rm F}\right)\qquad\mbox{otherwise},
\end{array}\right.
\end{equation}
where $\lambda_{\rm g}$ is the heliographic latitude of the box-centre. 
The tilt angles are positive if the leading polarity is nearer to the 
equator. Note that
these operations are done in a tangential plane through the box-centre.
Problems with measuring on a spherical surface are thus very small.
The tilt angles calculated are called `pseudo-tilt-angle' as 
in case of the Mt. Wilson data because magnetic polarity information is 
not available \citep{howard1991}. 

The polarity separation is then computed on the great circle (orthodrome) 
through the polarities:
\begin{equation}
\cos\Delta\beta =\sin\lambda_{\rm F}\sin\lambda_{\rm L}+\cos\lambda_{\rm F}\cos\lambda_{\rm L}\cos(\phi_{\rm F}-\phi_{\rm L}).
\end{equation}
The tilt angles are calculated for all spot groups with two or more
spots.

\begin{figure}
\includegraphics[width=0.49\textwidth]{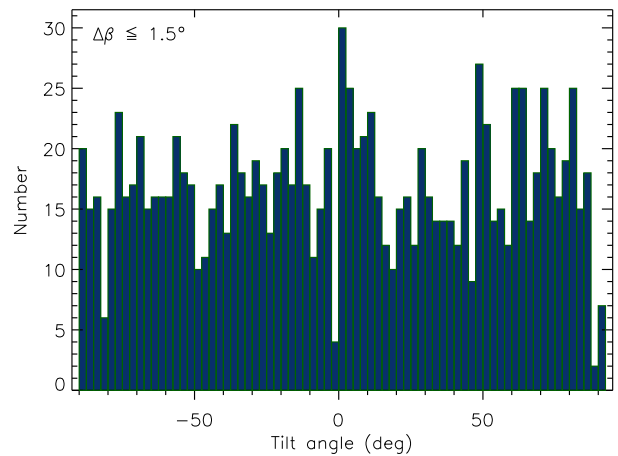}
\caption{Histogram of the `tilt angles' of supposedly unipolar
groups in the Schwabe data, selected by a maximum `polarity'
separation of $\Delta\beta\leq1\fdg5$.}
\label{tilthist_15}
\end{figure}

\subsection{Sources of errors\label{outliers}}
Differences between various data sources and tilt angle determinations
may have the following origins:
\begin{itemize}
\item unipolar groups are assigned a tilt angle erroneously,
\item the method of computation may introduce a bias if the
      division angle through the bipolar group is presumed or prejudiced,
\item ambiguity of the tilt angle sign due to the lack of magnetic information,
\item or incorrect splittings or combinations of groups lead
      to spurious tilt angles.
\end{itemize}
The misidentification of unipolar regions generates a noise component
whose distribution is much shallower than the distribution from bipolar
groups \citep{wang_ea2015,baranyi2015}. The peak-to-tail ratio apparent
from figure~8 in \citet{wang_ea2015} is about six for unipolar groups
with $\Delta\beta<2\fdg5$, while it is roughly 100 for the groups with
$\Delta\beta>2\fdg5$. In the Schwabe data, we find a peak-to-tail ratio
of a bit more than two for $\Delta\beta<2\fdg5$, while it is entirely
uniform for $\Delta\beta\leq1\fdg5$ (Fig.~\ref{tilthist_15}).
An interesting exercise is the determination of the dependence of the 
average tilt angle on the level of noise that is typically introduced
by unipolar sunspot groups that are erroneously included when no magnetic
information is available. We assume that the true tilt angle distribution
is symmetric around its mean and denote it by $S(\alpha,\alpha_0,
\sigma_\alpha)$, where $\alpha_0$ and $\sigma_\alpha$ are the true average 
and width of that distribution, respectively. We add the noise as a 
simple background constant $C$ which mimics the contamination by unipolar
groups or other spurious tilt angles to simplify the analysis, giving an upper limit
of the error introduced by those tilt angles. The average tilt angle is then
\begin{equation}
\langle\alpha\rangle=\int_{-\pi/2}^{\pi/2}\alpha\,(C+S)\,\mathrm{d}\alpha \,\left/\, 
  \int_{-\pi/2}^{\pi/2} (C+S)\, \mathrm{d}\alpha\right..
  \label{meanalpha}
\end{equation}
We can replace the constant $C$ by the number of sunspot groups that contribute 
to $C$ as a fraction of the total number of groups and denote this 
fraction by $f$. We find
\begin{equation}
 C = \frac{f\int_{-\pi/2}^{\pi/2} S \mathrm{d}\alpha}{\pi(1-f)}.
\end{equation}
Inserting this into (\ref{meanalpha}) leads to
\begin{eqnarray}
 \langle\alpha\rangle &=& (1-f)\int_{\pi/2}^{\pi/2}\alpha S \mathrm{d}\alpha \,\left/\,
 \int_{\pi/2}^{\pi/2} S \mathrm{d}\alpha\right.\nonumber\\
 &\approx&(1-f)\,\alpha_0.\label{alphaerror}
\end{eqnarray}
The approximation coming from the finite-limit integral is better 
than $0\fdg05$ up to $\alpha_0=42\degr$ for $\sigma_\alpha=20\degr$
and up to $15\degr$ for $\sigma_\alpha=30\degr$ -- good enough for
any relevant average tilt angles. The relation (\ref{alphaerror})
tells us that if 10\% of the individual 
tilt angles are spurious, the average tilt angle 
reduces by 10\%, e.g. from a true value of $5\degr$ to a 
measured value of $4\fdg5$. Looking at the cycle-to-cycle 
variations derived by \citet{wang_ea2015} for individual data 
sets, we infer cycle-to-cycle scatters (corrected by 
$t$-distribution) of 4\%, 9\%, and 14\% for the Debrecen 
umbral-based tilt angles, the Debrecen whole-spot-based 
tilt angles, and the tilt angles from the Mt.\ Wilson 
white-light images, respectively. Since the cycle-to-cycle 
variations of the average tilt angles are that small, the 
possible contamination of the distribution should be assessed,
especially when data from different sources are combined. Note
that 100\% noise naturally leads to an average tilt angle of
$0\degr$ (without polarity information). The influence of 
unipolar groups can be reduced significantly by excluding all
groups with apparent separations of $\Delta\beta<2\fdg5$ or
even $\Delta\beta<3\degr$ \citep{baranyi2015}.

\begin{figure} 
\includegraphics[width=0.49\textwidth]{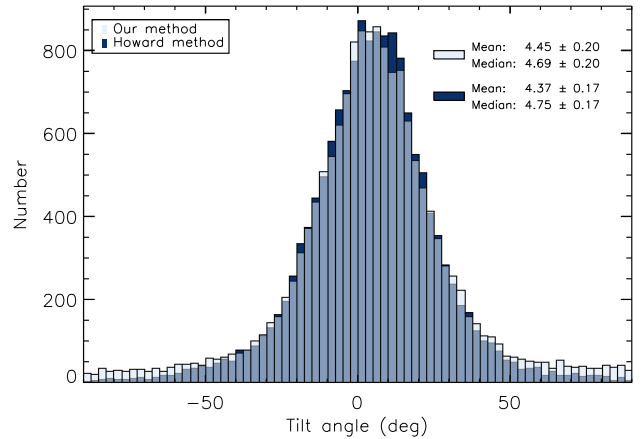}
\caption{Histograms of the tilt angles in the Schwabe
data, analysed with two different methods. Light bars show the
excess of tilt angles from our isotropic method searching for the
optimum polarity division, while dark bars show the excess of
tilt angles from the method by \citet{howard1991}. Only groups
with area weighted centres within $\pm60\degr$ CMD and polarity 
separations $\Delta\beta>3\degr$ are used.}
\label{tiltcomp}
\end{figure}

We tried to further reduce the influence of spurious tilt angles
by looking at the scatter of tilt angles of a given group during
its evolution over several days. We denote the individual appearances of
a group over several days as ``instances''. A similar procedure was proposed
by \citet{li_ulrich2012}. On the one hand, outliers due to ill-defined
groups need to be removed. On the other hand, groups become unipolar
at the end of their lifetime, but are still large and accompanied
by pores, mimicking $\Delta\beta>3\degr$. We therefore determine
the median tilt angle $\bar\alpha$ from the various instances of 
a given group and determine the average deviation from it by
\begin{equation}
  \Delta\alpha = \sum_{i=1}^{I}|\alpha_i - \bar\alpha|/I,
  \label{contamination}
\end{equation}
where $I$ is the number of instances of the group and $\alpha_i$
are the tilt angles of the individual instances of the group. The
tilt angle will be fairly reliable if the polarity separation is
large. We therefore tested whether the tilt angle at maximum polarity
separation, $\alpha_{\rm maxsep}$, does not deviate from the median 
significantly, using the criterion $|\alpha_{\rm maxsep}-\bar\alpha|<\Delta\alpha$. 
The groups fulfilling this criterion are a good guess
of the `real' bipolar groups, while others are omitted entirely.
Now, within the accepted groups, all instances with 
$|\alpha_i - \bar\alpha|>2\Delta\alpha$ are omitted as
outliers. A group turning unipolar near the end of its lifetime
still exhibits scattered spots and pores around the remaining 
(large-area) polarity which will cause quickly changing spurious
tilt-angles. Those (mostly H-type) groups are not supposed to deliver
a tilt angle. We will call those cases `evolutionary outliers' in
the following.

The removal of `evolutionary outliers' also requires the decision
on which hemisphere a given group lay, since low-latitude groups
may have instances on both sides of the equator, leading to jumps
in the tilt angle. We decided upon the hemispheric `membership' 
by the average (signed) latitude of the group instances of each group.
The signs of the tilt angles of all instances are then computed
assuming the single hemisphere obtained from that average latitude,
regardless of the actual hemisphere of an individual instance.

With respect to the method of computing tilt angles,
we used the Schwabe data set to compare the method by
\citet{howard_ea1984} with our method. The former divides groups
always with a north--south line leading to a bias with an avoidance of
tilt angles near $90\degr$. Our method of trying all possible dividing
lines is isotropic with equal prior probabilities for tilt angles
of $90\degr$, $0\degr$, and $-90\degr$. Figure~\ref{tiltcomp} shows a comparison
of the two methods based on the Schwabe data of individual spots.
The method of \citet{howard1991} tends to concentrate tilt angles
at lower values.

Groups which are `reversed dipoles' as compared to the typical 
polarity of a given cycle (anti-Hale groups) cannot be detected
in white-light images or sunspot drawings. The anti-Hale fraction
of all groups is about 8\% \citep{li_ulrich2012,mcclintock_ea2014}
or about 5\% \citep{sokoloff_khlystova2010} or even lower \citep{sokoloff_ea2015},
based on magnetogram data. This fraction holds true for large 
bipolar regions though, while the fraction may be as high as 50\%
for ephemeral regions with areas less than 50~MSH \citep{illarionov_ea2015},
which are not relevant here as they are not accompanied by sunspots.
In our cleanest distribution without `evolutionary outliers', 95\% 
of the groups are larger than 10~MSH in terms of their umbral 
group area. This translates to roughly 100~MSH total group area.
For this lower limit of group areas, \citet{illarionov_ea2015} give 
an anti-Hale fraction of about 37\%, while for large groups with
500~MSH or more, that fraction is 10\% or less.
The unsigned tilt angle distribution of anti-Hale groups is broader
than that of the Hale groups, but with similar peaks \citep{mcclintock_ea2014}.
The influence of the missing knowledge of the polarity on the average 
tilt angle is therefore relatively mild and not as strong as a 
similar fraction of random noise in the data.

The definition of what exactly is a group
yields another source of possible errors. 
\citet{baranyi2015} revisited the Mt.\ Wilson and Kodaikanal data sets and 
compared them with Debrecen tilt angles. Among other things, she found
that the automated routine used in the original analysis of the Mt.\ Wilson
and Kodaikanal data often splits true groups into two smaller ones. 
While the average tilt angle by
\citet{howard1991} ($4\fdg2\pm0\fdg2$) was reproduced as $4\fdg16\pm0.19$, 
a higher value of $4\fdg69\pm0\fdg20$ was found for 1917--1976 when these 
extra splittings were corrected. In terms of Eq.~\ref{contamination}, 
this new value indicates a random noise fraction in the original value of 11\%. That 
period of 1917--1976 is the one for which the Greenwich Photoheliographic 
Database was used as a reference to define proper groups. For 1974--1985, 
the Debrecen Photoheliographic Database was used to obtain the `true groups', 
to give the average of $5\fdg00\pm0\fdg47$. Note that the difference
to $4\fdg69\pm0\fdg20$ may actually be real and due to the different cycles covered.

\begin{figure} 
\includegraphics[width=0.49\textwidth]{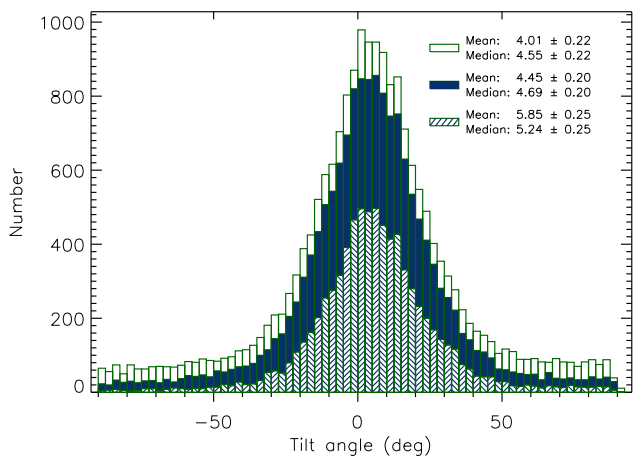}
\caption{Histograms of the tilt angles of groups with area weighted centres within $60\degr$ 
central meridian distance. {\em Open bars:} entire set, {\em filled bars:} groups 
with a minimum polarity separation of $\Delta\beta_{\rm min}=3\degr$,
and {\em hatched bars:} groups with $\Delta\beta_{\rm min}=3\degr$ and
a removal of `evolutionary' outliers that occur in the sequence of
tilt angles during the evolution of any given group. This hatched
histogram is an attempt to further reduce the influence of occasional
unipolar instances of otherwise bipolar groups. See text for the detailed
algorithm. The bin width is $2\fdg5$.}
\label{tilthist}
\end{figure}

\begin{figure} 
\includegraphics[width=0.49\textwidth]{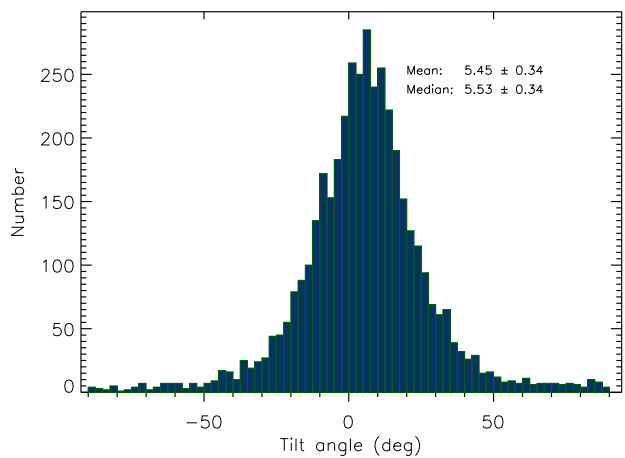}
\caption{Tilt angle histogram of the Schwabe data for group centres within 
$60\degr$ central meridian distance and polarity separations $\Delta\beta>3\degr$.  
In this analysis, only spots with umbral areas of 5~MSH or more were used
to compute the weighted positions of the polarities. The selection therefore
affects both tilt angles and polarity separations. As compared with the
hatched bars in Fig.~\ref{tilthist}, many groups have now turned into
unipolar groups, since one of the polarities was represented by a 
single spot of less than 5~MSH.}
\label{tilt5msh}
\end{figure}

\subsection{Distribution and averages\label{averages}}
Figure~\ref{tilthist} shows the resulting distribution of tilt angles. Being
quite broad, the distribution has its maximum at small, non-zero $\alpha$. 
Note that the open bars may include tilt angles which are erroneously computed for two or more spots of 
a single polarity, fragmented spots, and spots inside the same penumbra.
The filled histogram therefore shows the distribution of tilt angles with 
polarity separations $\Delta\beta>3\degr$. In this distribution, spots inside 
a common penumbra are essentially excluded, but, on the one hand, true 
bipolar groups with very small polarity separations may also be 
excluded. On the other hand, spurious tilt angles due to a 
decaying big group with a single polarity
may still contribute to this distribution, but are a very minor 
fraction \citep{baranyi2015}. The selection of bipolar
groups may perhaps be fine-tuned by using an area-dependent minimum polarity
separation, but we did not wish to impose biases which may 
affect the distribution of polarity separations. A constant minimum
separation $\Delta\beta_{\rm min}$ is therefore used for selecting 
actual bipolar groups.

The average tilt angle for 
the distribution with $\Delta\beta_{\rm min}=3\degr$ and with area weighted
group centres within $\pm60\degr$ CMD is $4\fdg45\pm0\fdg20$ 
(median $4\fdg69\pm0\fdg20$) where the error of the mean is computed by the 
standard deviation of the distribution divided by the square root of the number 
of points, $\sigma_{\rm tilt}/\sqrt{n}$. For comparison, we may also 
compute the average tilt angle for $\Delta\beta_{\rm min}=0$
(again $|{\rm CMD|}<60\degr$) and obtain $4\fdg01\pm0\fdg22$ (median $4\fdg55\pm0\fdg22$). 
The lower value is consistent with our earlier supposition that 
spurious bipolarities add a certain amount of randomness to the 
data, bringing the average closer to zero. This average agrees relatively well with 
the ones found by \citet{howard1991} ($4\fdg2\pm0\fdg2$) and \citet{dasi_ea2010}
($4\fdg25\pm0\fdg18$ for Mt. Wilson and $4\fdg51\pm0\fdg18$ for Kodaikanal) for
solar cycles~15--21 which were all computed without a lower limit, i.e.\ $\Delta\beta_{\rm min}=0$.

An analysis of the Debrecen data by \citet{baranyi2015} with careful
extraction of truly bipolar groups delivered $5\fdg12\pm0\fdg46$ for 1974--1985
(end of cycle~20 and cycle~21). Based on a minimum polarity separation of $\Delta\beta_{\rm min}=3\degr$,
we recomputed the Mt.\ Wilson and Kodaikanal averages as well which resulted in
values of $5\fdg95\pm0\fdg42$ and $6\fdg91\pm0\fdg45$, respectively, for 
cycle~21. \citet{ivanov2012} used the Pulkovo database (Catalogue of Solar Activity 
-- CSA) for tilt angles in the period 1948--1991. We used their database and
obtained an average tilt angle of $6\fdg41\pm0\fdg14$ for cycle~21 only. 
In this sample, the groups were determined manually and 
contain a fairly clean definition of what a group is, similar to our analysis of 
the Schwabe drawings. Since the average tilt angle varies from one cycle
to the next, we cannot compare Schwabe's tilt angles with 20th-century ones 
directly, but we find that averages of clean tilt angle samples are typically 
$5\degr$ or larger.

\begin{figure} 
\includegraphics[width=0.49\textwidth]{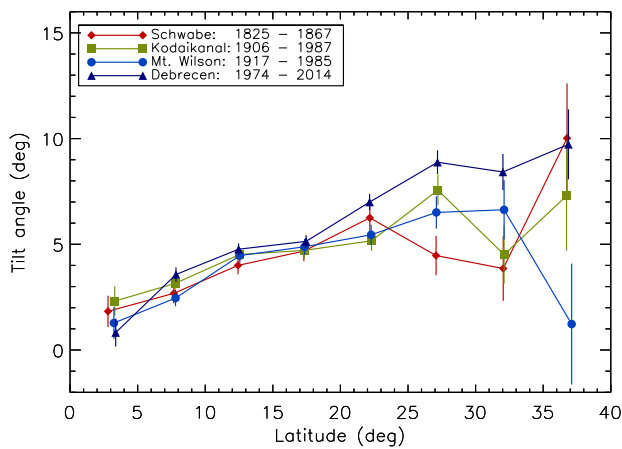}
\caption{Dependence of the average tilt angle on the unsigned latitude
for the data from Schwabe, Kodaikanal, Mt. Wilson, and Debrecen. Only
groups with central meridian distances within $\pm60\degr$ are used as 
the simplest common filter for all datasets.
Errors of the means are the standard deviation of the bin sample 
divided by $\sqrt{n}$ in each bin.}
\label{joyslaw}
\end{figure}



The resulting histogram of tilt angles with deleted `evolutionary 
outliers' (Sect.~\ref{outliers}) is shown in Fig.~\ref{tilthist} 
as hatched bars. The average tilt angle has risen to 
$5\fdg85\pm0\fdg25$, based on 7765~tilt angles. We consider this 
the cleanest sample of tilt angles for cycles~7--10.
The only average this value can be compared with now 
(as far as different cycles can be compared at all) is the 
one given by \cite{baranyi2015} where bipolar groups have 
also been selected fairly rigorously. However, that Debrecen 
tilt angle of $5\fdg12$ is different in that spots smaller 
than 5~MSH are considered as pores which do not enter the determination of tilt 
angles and polarity separations. Figure~\ref{tilt5msh} shows the
histogram of the tilt angles restricted to spots that have areas $\geq 5$~MSH.
The average of $5\fdg45\pm0\fdg34$ (based on 4154 tilt angles) is slightly lower
than the above value of $5\fdg85\pm0\fdg25$, but not significantly. Even though
the pores seem to play a minor role in determining reliable tilt angles, the
difference shows that the minimum umbral area needs to be considered when combining 
several data sets.

\begin{table*}
\caption{Modified format of the data of individual sunspots observed by
Samuel Heinrich Schwabe. The format extends the one by \citet{arlt_ea2013} after column 93.
In the Format column, I denotes integer fields, C8 is an 8-character text field, and, e.g., F5.1
denotes a 5-character-wide floating point field with one decimal. Areas in UMB are based on the
derivation by (\ref{mapping}).\label{format}}
\begin{tabular}{lllp{12.9cm}}
\hline
\hline
Field & Column & Format & Explanation \\
\hline
{\tt YYYY}  & 1--4   & I4     & Year \\
{\tt MM}    & 6--7   & I2     & Month \\
{\tt DD}    & 9--10  & I2     & Day referring to the German civil calendar running from midnight to midnight\\
{\tt HH}    & 12--13 & I2     & Hour, times are mean local time in Dessau, Germany\\
{\tt MI}    & 15--16 & I2     & Minute, typically accurate to 15~minutes\\
{\tt T}     & 18     & I1   & Indicates how accurate the time is. Timeflag\,=\,0 means the time has
                 been inferred by the measurer (in most cases to be 12h~local time);
                 Timeflag\,=\,1 means the time is as given by the observer\\
{\tt L0}    & 20--24 & F5.1   & Heliographic longitude of apparent disk centre seen from Dessau\\
{\tt B0}    & 26--30 & F5.1   & Heliographic latitude of apparent disk centre seen from Dessau\\
{\tt CMD}   & 32--36 & F5.1   & Central meridian distance, difference in longitude from disk centre;
                 contains --.- if line indicates spotless day;
                 contains NaN if position of spot could not be measured.\\
{\tt LLL.L} & 38--42 &F5.1 & Heliographic longitude in the Carrington rotation frame;
                 contains --.- if line indicates spotless day;
                 contains NaN if position of spot could not be measured.\\
{\tt BBB.B} & 44--48 & F5.1 & Heliographic latitude, southern latitudes are negative;
                 contains --.- if line indicates spotless day;
                 contains NaN if position of spot could not be measured.\\
{\tt M}     & 50    & C1   & Method of determining the orientation. `C': horizontal pencil line parallel to
                 celestial equator; `H': book aligned with azimuth-elevation; `Q': rotational
                 matching with other drawings (spot used for the matching have ${\rm ModelLong}\neq {\rm `-.-'}$,
                 ${\rm ModelLat}\neq {\rm `-.-'}$, and ${\rm Sigma}\neq {\rm `-.-'}$).\\
{\tt Q}     & 52    & I1   & Subjective quality, all observations with coordinate system drawn by Schwabe get
                 Quality\,=\,1.
                 Positions derived from rotational matching may also obtain Quality\,=\,2 or 3, if the
                 probability distributions fixing the position angle of the drawing were not
                 very sharp, or broad and asymmetric, respectively. Spotless days have Quality\,=\,0;
                 spots for which no position could be derived, but which have sizes, get Quality\,=\,4.\\
{\tt SS}       & 54--55  & I2   & Size estimate in 12 classes running from 1 to 12; a spotless day is indicated by 0 \\
{\tt GROUP}    & 57--64  & C8   & Group designation based on Schwabe, but modified by our regrouping\\
{\tt MEASURER} & 66--75  & C10  & Last name of person who obtained position\\
{\tt MOD\_L}   & 77--81  & F5.1 & Model longitude from rotational matching (only spots used for the matching have this)\\
{\tt MOD\_B}   & 83--87  & F5.1 & Model latitude from rotational matching (only spots used for the matching have this)\\
{\tt SIGMA}    & 89--93  & F6.3 & Total residual  of model positions compared with measurements of reference spots in rotational matching (only spots used for the matching have this). Holds for entire day.\\
{\tt DELTA}    & 96--99  & F4.1 & Heliocentric angle between the spot and the apparent disk centre in degrees (disk-centre distance); it is $-.-$ for spotless days, while it is NaN if the spot position could not be determined.\\
{\tt UMB}      &100--103 & I3   & Inferred umbral area in millionths of the solar hemisphere (MSH); it is 0 for spotless days and NaN if spot position could not be derived or ${\tt DELTA}>85\degr$\\
{\tt A}        &105      & C1   & Flag saying whether area mapping is based on umbral (`U') or penumbral (`!') areas with the latter being less certain. Note that the actual area given in {\tt UMB} is always umbral. Spotless days have $-$.\\
\hline
\end{tabular}
\end{table*}

\begin{table*}
\caption{Format for the tilt angle data derived from the sunspot groups observed by Schwabe,
with format symbols as in Table~\ref{format}.
All areas are based on the UMB column in Table~\ref{format}.}
\begin{tabular}{lr@{--}llp{13.4cm}}
\hline
\hline
Field  & \multicolumn{2}{c}{Column} & Format & Explanation \\
\hline
{\tt YYYY}   & 1&4    & I4     & Year\\
{\tt MM}     & 6&7    & I2     & Month\\
{\tt DD}     & 9&10   & I2     & Day\\
{\tt HH}     & 12&13  & I2     & Hour\\
{\tt MI}     & 15&16  & I2     & Minute; mean local time in Dessau, Germany\\
{\tt GROUP}  & 18&25  & C8     & Group name based on Schwabe, but modified by our regrouping\\
{\tt SP}     & 27&28  & I2     & Number of spots in a group\\
{\tt ARA}    & 30&32  & I3     & Sum of umbral area of all spots in a group, in millionths of the solar hemisphere (MSH)\\
{\tt AWL.L}  & 34&38  & F5.1   & Area-weighted heliographic longitude of the group\\
{\tt AWB.B}  & 40&44  & F5.1   & Area-weighted heliographic latitude of the group\\
{\tt TILTAN} & 46&51  & F6.2   & Tilt angle of the group; positive sign means leading polarity closer to equator in either hemisphere. This
                                  tilt angle was found using an isotropic search for the most likely dividing line between the polarities.\\
{\tt TILTHO} & 53&58  & F6.2   & Tilt angle computed as in \citet{howard1991} for compatibility reasons. It is based on a fixed 
                                  vertical dividing line between the polarities and an approximative formula for the tilt angle.\\
{\tt POLSP}  & 60&64  & F5.2   & Polarity separation of the group in degrees on the solar sphere. This and the following items are based on the polarity definition for {\tt TILTAN}.\\
{\tt FN}     & 66&67  & I2     & Number of spots in the following polarity\\
{\tt LN}     & 69&70  & I2     & Number of spots in the leading polarity\\
{\tt FAR}    & 72&74  & I3     & Umbral area of the following polarity, in MSH\\
{\tt LAR}    & 76&78  & I3     & Umbral area of the leading polarity, in MSH\\
{\tt FLL.L}  & 80&84  & F5.1   & Area-weighted longitude of the following polarity\\
{\tt FBB.B}  & 86&90  & F5.1   & Area-weighted latitude of the following polarity\\
{\tt LLL.L}  & 92&96 &  F5.1   & Area-weighted longitude of the leading polarity\\
{\tt LBB.B}  & 98&102 & F5.1   & Area-weighted latitude of the leading polarity\\
{\tt GFC}    &104&108 & F5.1   & Heliocentric distance of the group from the disk centre in degrees\\
\hline
\label{tilt_tab_format}
\end{tabular}
\end{table*}

The average dependence of the tilt angle on the absolute heliographic
latitude (Joy's law) is shown in Fig.~\ref{joyslaw}. Together with the
Schwabe data, we also plotted the average tilt angles obtained from
the Mt.~Wilson, Kodaikanal, and Debrecen data, using only tilt angles 
from groups within $\pm60\degr$ CMD. The 
latitudinal dependence in the Schwabe data may be a bit shallower
than the dependences from the other data sets -- especially since
it exhibits a non-zero intersection with the ordinate --, but is in
agreement with them considering the uncertainty margins. 

The format of the database containing the individual
umbrae including umbral areas as well as spotless days is given in 
Table~\ref{format} which extends the one given by \cite{arlt_ea2013}, while the format of the final data set containing the 
tilt angles and polarity separations of sunspot groups is given in Table~\ref{tilt_tab_format}.
Note that we give both, the tilt angles obtained according to our method 
described above as well as the tilt angles derived with the method
by \citet{howard1991} in the latter table.


\section{Summary}\label{results}
This study aims at determining physical areas of sunspots from drawings
by Samuel Heinrich Schwabe in 1825--1867 as well as ordering these sunspots
into (hopefully bipolar) groups and computing tilt angles
of these sunspot groups for that period. The fraction of the solar disk
covered by the pencil dots in the drawings cannot be directly converted
into an area in km$^2$ or millionths of a solar hemisphere (MSH). We 
therefore constructed a mapping of the twelve arbitrary cursor sizes which 
were used by \citet{arlt_ea2013} to estimate the sizes of sunspots in the 
Schwabe drawings. For cycles~8--10, we obtain an average umbral area 
per day of 113. The Debrecen data for cycles~21--23, which were predominantly 
used for calibration, yield an average of roughly 150. The difference appears
to be compatible with the stronger cycles in the second half of the
20th century and the fact, that Schwabe may have overlooked (or not
plotted) a number of small spots. The umbral areas in the Greenwich 
Photoheliographic Database lead to an average of about 140 from
cycles~12--20 of mixed strengths. Our area conversion is independent 
of the Greenwich data, but seems to agree with it fairly well, again 
taking into account that the Schwabe drawings may miss a few smaller 
spots.

The area distribution of the Schwabe sunspots exhibits a log-normal 
distribution in agreement with 20th-century data \citep{bogdan_ea1988} and is essentially independent of 
the cycle phase. Schwabe's original sunspot group designations were 
modified so that the groups comply with the modern understanding of a 
sunspot group, with the limitation of missing magnetic information. 
Using the positions and areas of all individual sunspots, the tilt 
angles as well as the polarity separations of the sunspot groups were 
calculated. Without the magnetic information, the definition of the 
polarities may lead to wrong associations affecting both tilt angles 
and polarity separations. The manual inspection of the groups before 
computing these quantities reduces these incorrect polarities as compared 
to fully automatic analysis schemes. Nevertheless, a remaining random
component in the tilt angle distribution is likely to be present. 

Both an updated sunspot database and a tilt angle database are available at 
http://www.aip.de/Members/rarlt/sunspots for further study. We note that
in the sunspot database:
\begin{itemize}
\item sunspot areas for spots with ${\rm CMD}\leq 70\degr$ are reliable,
\item sunspot areas for spots with $70\degr<{\rm CMD}\leq85\degr$ are
uncertain because they are an extrapolation of the statistical method employed,
\item sunspot areas for spots with ${\rm CMD}>85\degr$ have been omitted,
\item sunspot areas are calibrated using 20th century data; they will
not serve for the purpose of detecting differences in areas 
between the 19th and the 20th century.
\end{itemize}
In the tilt angle database:
\begin{itemize}
\item tilt angles for all groups with two or more spots are reported,
\item tilt angles for groups with ${\rm CMD}\leq60\degr$ are considered
reliable as the positions are reliable,
\item tilt angles for polarity separations $\Delta\beta>3\degr$ (POLSP) are 
likely to be bipolar groups and should be selected for further analysis.
\item the influence of spurious tilt angles from remaining unipolar groups
can be further reduced by removing outliers from the sequence of tilt angles
provided by the evolution of a given group. In brief, we removed days
of appearance of a given group if the obtained tilt angle deviates 
significantly from the mean tilt angle of all appearances of that single 
group. The actual procedure is a bit more involved and is described in 
Sect.~\ref{outliers}.
\end{itemize}
Joy's law was found to be obeyed by the likely bipolar groups.
The latitude dependence averages over all Schwabe cycles is not 
significantly different from cycles in the 20th century.

The applicability of white-light images for inferring cycle
properties was doubted by \citet{wang_ea2015}, mostly because
of the inevitable contamination by actual unipolar groups. 
We believe the white-light images and drawings can still be
useful, because (a) improved algorithms and visual inspection
can reduce the impact of unipolar groups significantly, and
because (b) we are interested in relative variations of tilt
angles from one cycle to another, so consistently analysed 
sunspot data can still provide useful relative information
of cycle-to-cycle variability. Care has to be taken that all 
data used for a particular study are consistent with each other,
e.g.\ tilt angles from white-light images and magnetograms
cannot be combined directly into a single record.

We have not studied the tilt angles of the individual
cycles in this Paper. This will be the subject of a future
study of tilt angles and strengths of invidual cycles
as well as correlations thereof, extending ealier works 
on Mt.\ Wilson and Kodaikanal data.  Especially cycle~7 
is an interesting candidate for peculiarities, since it 
occurred shortly after the Dalton minimum (roughly 1795--1820). 
The current Paper aims at the disemination of the areas and 
tilt angles as the basis for further studies.


\begin{acknowledgements}
We would like to thank Robert Cameron for helpful discussions
and Julian Kern for inspecting the spots in Schwabe's magnified drawings.
This study was supported by the German \emph{Deut\-sche 
For\-schungs\-ge\-mein\-schaft, DFG\/} project number Ar~355/10-1.
It was partly supported by the BK21 plus program through 
the National Research Foundation (NRF) funded by the Ministry of 
Education of Korea.
\end{acknowledgements}


\bibliographystyle{aa}
\bibliography{schwabe_tilt}

\end{document}